\documentclass[prd,aps,nofootinbib,notitlepage,showpacs,showkeys,preprintnumbers]{revtex4-1}
\usepackage{graphicx,epsf,amsmath,amsfonts,amssymb,amsbsy,enumerate}
\usepackage{graphicx}
\usepackage{epsfig}
\usepackage{epstopdf}

\graphicspath{{Figures/}}

\newcommand{\ds}{\displaystyle}
\newcommand{\vev}[1]{\langle#1\rangle}
\newcommand{\mat}{\left ( \begin{array}}
\newcommand{\emat}{\end{array} \right )}
\newcommand{\vect}{\left ( \begin{array}{c}}
\newcommand{\evect}{\end{array} \right )}

\newcommand{\Det}{\mathop{\rm Det}\nolimits}

\begin{document}

\title{ \bf Chiral imbalanced hot and dense quark matter: NJL analysis at the physical point and comparison with lattice QCD
}

\author{T. G. Khunjua $^{1)}$, K. G. Klimenko $^{2)}$, and R. N. Zhokhov $^{3)}$}

\vspace{1cm}

\affiliation{$^{1)}$ Faculty of Physics, Moscow State University, 119991, Moscow, Russia}
\affiliation{$^{2)}$ State Research Center of Russian Federation -- Institute for High Energy Physics,
NRC "Kurchatov Institute", 142281, Protvino, Moscow Region, Russia}
\affiliation{$^{3)}$ 
Pushkov Institute of Terrestrial Magnetism, Ionosphere and Radiowave Propagation (IZMIRAN),
108840 Troitsk, Moscow, Russia}

\begin{abstract}
 Hot and dense quark matter with isospin and chiral imbalances is 
investigated in the framework of the (3+1)-dimensional Nambu--Jona-Lasinio model (NJL) in the large-$N_c$ limit ($N_c$ is the number of quark colors). Its phase structure is considered in terms of barion -- $\mu_B$, isospin -- $\mu_I$ and chiral isospin -- $\mu_{I5}$ chemical potentials. It is shown in the paper that (i) in the chiral limit there is a duality between chiral symmetry breaking (CSB) and charged pion condensation (PC) phenomena. (ii) At the physical point, i.e. at nonzero bare quark mass $m_0$, and temperature this duality relation is only approximate, although rather accurate. (iii) We have shown that the chiral isospin chemical potential $\mu_{I5}$ in dense quark matter generates charged pion condensation both at zero and nonzero $m_0$, and at $\mu_{I5}\ne 0$ this phase might be observed up to  temperatures as high as 100 MeV. (iv) Pseudo-critical temperature of the chiral crossover transition rises in the NJL model with increasing $\mu_{I5}$.
(v) It has been found an agreement between particular sections of the phase diagram in the framework of NJL model and corresponding ones in lattice QCD simulations. Two different plots from different lattice simulations that are completely independent and are not connected at the first sight are in reality dual to each other, it means that lattice QCD simulations support the hypothesis that in real quark matter there exists the (approximate) duality between CSB and charged PC. Moreover, we can reverse the logic and we can predict the increase of pseudo-critical temperature with 
chiral chemical potential, the much debated effect recently, just by the duality notion, hence bolster 
 confidence in this result (lattice QCD showed this feature for unphysically large pion mass) and put it on the considerably more solid ground.

\end{abstract}


\maketitle

\section{Introduction}

At normal (Earth) conditions, protons and neutrons form atomic nuclei, and the latter, together with their orbital electrons, form the ordinary matter of our environment. If matter is subjected to extreme compression, eventually all chemical and nuclear bonds are broken, and the matter is squeezed from the molecular scale to the sub-particle scale with density higher than 0.15 baryon per $\rm {fm^3}$. Experimental creation of such dense matter is a very hard problem but such conditions can take place inside compact stars due to compression by gravity into a stable and extremely dense state.
As a rule neutron stars have comparatively low temperatures and one can assume that it is zero. What happens at high temperature is hard to probe studying the physics of neutron stars. Nevertheless, due to technology advances, modern accelerators of elementary particles are now able to collide not only single high energy protons, but also heavy ions consisting of many coupled protons and neutrons.  It is believed that in the fireball just after heavy-ion collisions there emerges a droplet of quark gluon plasma with very high temperature. Physics of heavy ion collision experiments can shed some light on the conditions that existed a few microseconds after the Big Bang and provide answers to several other questions.

The fundamental theory of matter in such extreme conditions is quantum chromodynamics (QCD) which is a gauge field theory associated with $SU(3)$ group, where gauge bosons (gluons) play the role of interaction carriers of quarks. 
The main method of QCD analysis is the perturbative technique on the basis of coupling constant. However, it is not always possible to use this technique, as QCD calculations can be too complex or in the low energy region when the coupling constant is too large.
In particular, QCD perturbative technique is not applicable in a consideration of physically reachable dense matter
, etc.
In these cases, non-perturbative methods, such as effective theories or lattice calculations, are usually used.

Thus, the entire QCD phase diagram could not be described currently in the framework of a unified theory. Lattice calculations are very useful for description of the region of zero density and high temperature. However, the so-called sign problem still presents insurmountable difficulties for lattice calculations in the nonzero density region. On the other hand effective theories do not have fundamental background and as a result do not share the main prominent features with QCD such as a gauge invariance, renormalizability, etc. Nevertheless, at this moment, effective models are the best tool for investigating dense quark matter. At this time one of the most widely used effective model is the Nambu--Jona-Lasinio (NJL) model \cite{Nambu:1961fr,Klevansky:1992qe,Hatsuda:1994pi,Buballa:2003qv}. 

It is well known that usually 
dense baryonic matter in compact stars obeys an isospin asymmetry, i.e. where the densities of up- and down quarks are different (it is characterised by isospin chemical potential $\mu_I$). In experiments on heavy-ion collisions, we also have to deal with quark matter which has an evident isospin asymmetry because of different neutron and proton contents of colliding ions. In early 70-th Sawyer \cite{Sawyer:1972cq} and independently Migdal \cite{Migdal:1973} have shown that there might be phase transition from pure neutron matter to mixed hadron matter with protons, neutrons and $\pi^0$-pions at superdense matter in the compact stars. Later, using the chiral perturbation theory, it was shown that there is a threshold $\mu_I^c=m_\pi\approx 140 MeV$ of a phase transition to the charged pion condensation (PC) phase \cite{Son:2000}. This result was ultimately proved in the framework of random matrix model \cite{Klein:2003}, Ladder-QCD model \cite{Barducci}, resonance gas model \cite{Toublan}, quark-meson model \cite{Fraga}, NJL model \cite{ek1} (including (1+1)-dimensional version of the NJL model \cite{ek2}) and lattice simulations \cite{Gupta}. Nevertheless, the whole picture is still a matter of debate.

Now the main question is whether the charged pion condensation exists in the real world and how this phenomenon behaves under influence of various external factors. And different factors can have a completely different effect on this phase. For example, in the framework of NJL model the finite-size effects, spatial inhomogeneity of the pion condensate \cite{Khunjua} or chromomagnetic background field \cite{Fedotov} could promote the charged PC phase. On the other hand,
if the electric charge neutrality and $\beta$-equilibrium constraints are imposed,  the charged PC phenomenon in quark matter depends strongly on the bare (current) quark mass values. In particular, it turns out that the charged PC phase with {\it nonzero baryonic density} is not realized within NJL models, if the bare quark mass $m_0$ reaches the physically acceptable values of $5\div 10$ MeV \cite{abuki}, i.e. at the physical point. In addition, temperature $T$ and different model parameters such as coupling constants, etc, as well strongly influence on this phase \cite{Ebert}. It is also worth to note that the phase structure of the isospin imbalanced quark matter below the threshold ($\mu_I<m_\pi$) is an important question because even small nonzero $\mu_I$ could double the critical endpoint of a phase diagram and affects the results of heavy-ion collision experiments \cite{Klein:2003,Kogut}.

Recently, it has been shown in the framework of the massless (3+1)-dimensional NJL model (and in the leading large-$N_c$ order, where $N_c$ is the number of colors of quarks) that chiral imbalance promotes charged PC phase in dense matter at zero temperature \cite{Khunjua:2017mkc,Khunjua:2018sro} and responsible for the existence of the duality between chiral symmetry breaking (CSB) and charged PC phases. The imbalance between densities of left-handed and right-handed quarks (chiral imbalance) is a highly anticipated phenomenon that could occur both in compact stars and heavy ion collisions. This effect could stem from nontrivial interplay of axial anomaly and the topology of gluon configurations.\footnote{It is predicted that there is an electrical current in the chiral imbalanced quark matter under strong magnetic field \cite{Fukushima:2008xe}. This phenomenon is called chiral magnetic effect and it could be an evidence of the chiral imbalance in QCD.} Also, there is another mechanism of its origin -- chiral separation effect which can be realized in dense matter in the presence of a strong magnetic field. In this case left-handed and right-handed quarks tend to move in opposite directions along the magnetic field, thereby creating regions with chiral imbalance. Moreover, in the case of two-flavored quark matter the chiral separation effect could promote (see below in Appendix A) both nonzero chiral density $n_5$ and nonzero isotopic chiral density $n_{I5}$, and quark matter can be described using the corresponding chemical potentials $\mu_5$ and $\mu_{I5}$.

It was already mentioned above that nonzero bare quark mass $m_0$ and nonzero temperature $T$ could destroy charged PC phase in the physically adequate circumstances. So one of the aims of our present work is to check the robustness of the charged PC phase generated by chiral imbalance under the influence of these destructive factors. Another purpose is to study in the framework of the NJL$_4$ model the fate of the duality observed in the chiral limit \cite{Khunjua:2017mkc} (where it is an exact symmetry) between CSB and charged PC phenomena in the leading large-$N_c$ order: we investigate the influence of the bare quark mass and temperature on this effect, etc. In particular, it is shown in our paper that duality correspondence between CSB and charged PC still is a very good approximate symmetry of a phase portrait of the NJL$_4$ model even at $m_0\ne 0$ and $T\ne 0$.

It is interesting to investigate not only the charged PC phase but also hot quark matter itself with chiral asymmetry only. In this case  at zero baryon chemical potential, $\mu_B=0$, there is no sign-problem and we have solid results from lattice simulations \cite{Braguta:2015zta,Braguta:2015owi}. Nevertheless, some key properties of chirally imbalanced quark matter are still under debate. So, in addition to charged PC phase, in the present paper we also investigate in the framework of the NJL$_4$ model at $m_0\ne 0$ the dependence of the (pseudo-)critical temperature, which characterizes the chiral cross-over region of the phase diagram, on the chiral isospin chemical potential $\mu_{I5}$ and compare our results with other effective model investigations and lattice simulations on this topic. Note also that at $\mu_B=0$ and $m_0\ne 0$ the $(\mu_I,T)$- and $(\mu_5,T)$-phase diagrams  have been obtained both using lattice QCD simulations and in the framework of the NJL model, and the results are in good agreement. Moreover, in the present paper we show that just these phase diagrams are dually conjugated (with a good precision) to each other, so there is a good reason to argue that duality between CSB and charged PC phenomena is confirmed by lattice QCD calculations.

The paper is organized as follows. In Sec. II a (3+1)-dimensional NJL model with two massive quark flavors ($u$ and $d$ quarks) that includes three kinds of chemical potentials, $\mu_B,\mu_I,\mu_{I5}$, is introduced. Furthermore, the symmetries of the model are discussed and its thermodynamic potential is presented in the leading order of the large-$N_c$ expansion both at zero and nonzero temperature $T$. In particular, it is shown in this section that in the chiral limit ($m_0=0$) the phase structure of the model (in the leading order over $1/N_c$) has a dual symmetry between CSB and charged PC phenomena.
In the next section we formulate the main consequences of the exact dual symmetry (Sec. III A), using which it is possible to decide that dual symmetry is performed approximately in the NJL model at $m_0\ne 0$ and $T=0$, but with good accuracy (Sec. III B). It Sec. III C we show that at nonzero values of the chiral isospin chemical potential $\nu_{I5}$ the charged PC phase with nonzero quark density can be realized in the model up to rather high values of temperature, $T\approx 100$ MeV. Moreover, here we show that duality is also fulfilled approximately at $T\ne 0$. In Sec. III D the plot of the pseudo-critical temperature of the chiral crossover transition as a function of $\mu_{I5}$ at $\mu=\mu_I=\mu_5=0$ is obtained. Here it is compared with results of other effective models and lattice QCD approaches. Sec. IV presents summary and discussion leading to the conclusion that duality between CSB and charged PC observed in the NJL$_4$ model is supported by some phase diagrams obtained by lattice QCD simulations at $\mu_B=0$.  Some technical details and issues not directly related to this work are relegated to Appendices A and B.

\section{The model and its thermodynamic potential}

\subsection{Lagrangian and symmetries}

It is well known that in the framework of effective four-fermion field theories dense and isotopically asymmetric quark matter, composed of $u$ and $d$ quarks, can be described by the following (3+1)-dimensional NJL Lagrangian
\begin{eqnarray}
&&  L=\bar q\Big [\gamma^\nu\mathrm{i}\partial_\nu -m_0
+\frac{\mu_B}{3}\gamma^0+\frac{\mu_I}2 \tau_3\gamma^0\Big ]q+ \frac
{G}{N_c}\Big [(\bar qq)^2+(\bar q\mathrm{i}\gamma^5\vec\tau q)^2 \Big
]. \label{1}
\end{eqnarray}
Here $q$ is a flavor doublet, $q=(q_u,q_d)^T$, where $q_u$ and $q_d$ are four-component Dirac spinors as well as color $N_c$-plets of the $u$ and $d$ quark fields, respectively (the summation in Eq. (\ref{1}) over flavor, color, and spinor indices is implied); $\tau_k$ ($k=1,2,3$) are Pauli matrices; $m_0$ is the bare quark mass (for simplicity, we assume that $u$ and $d$ quarks have the same mass); $\mu_B$ and $\mu_I$ are chemical potentials which are
introduced in order to study quark matter with nonzero baryon and isospin densities, respectively.

The symmetries of the Lagrangian (\ref{1}) depends essentially on wether the bare quark mass $m_0$ and chemical potentials take zero or nonzero values. For example, in the most particular case, when $m_0=\mu_{I}=0$ the Lagrangian (\ref{1}) is invariant under transformations from chiral $SU(2)_{L}\times SU(2)_{R}$ group, which is also inherent in 2-flavor QCD in the chiral limit. This symmetry is reduced to $U_B(1)\times U_{I_3}(1)\times U_{AI_3}(1)$ group if all chemical potentials are nonzero, and $m_0=0$. In this case the abelian baryon $U_B(1)$, isospin $U_{I_3}(1)$ and chiral isospin $U_{AI_3}(1)$ subgroups act on flavor doublet $q$ in the following way 
\begin{eqnarray}
U_B(1):~q\to\exp (\mathrm{i}\alpha/3) q;~
U_{I_3}(1):~q\to\exp (\mathrm{i}\alpha\tau_3/2) q;~
U_{AI_3}(1):~q\to\exp (\mathrm{i}
\alpha\gamma^5\tau_3/2) q.
\label{2001}
\end{eqnarray}
As a result, we see that in the chiral limit ($m_0=0$) the quantities $\hat n_B\equiv\bar q\gamma^0q/3$, $\hat n_I\equiv\bar q\gamma^0\tau^3 q/2$ and $\hat n_{I5}\equiv\bar q\gamma^0\gamma^5\tau^3 q/2$ are the density operators of the conserved baryon, isospin and chiral isospin charges of the system (1), respectively. Introducing the particle density operators for $u$ and $d$ quarks, $\hat n_u\equiv q_u\gamma^0q_u$ and $\hat n_d\equiv q_d\gamma^0q_d$, we have
\begin{eqnarray}
\hat n_B=\frac 13\left (\hat n_u+\hat n_d\right ),~~\hat n_I=\frac 12\left (\hat n_u-\hat n_d\right ).
\label{2003}
\end{eqnarray}
One can also introduce the particle density operators $\hat n_{fR}$ and  $\hat n_{fL}$ for right- and left-handed quarks of each flavor $f=u,d$ (see in Appendix \ref{ApB}). In this case the density operator of the chiral isospin charge looks like
\begin{eqnarray}
\hat n_{I5}=\frac 12\left (\hat n_{uR}-\hat n_{uL}-\hat n_{dR}+\hat n_{dL}\right )=\frac 12\left (\hat n_{u5}-\hat n_{d5}\right ),
\label{2004}
\end{eqnarray}
where the quantity $\hat n_{f5}\equiv \hat n_{fR}-\hat n_{fL}$ is usually called the density operator of the chiral charge for the quark flavor $f=u,d$. Below, in Appendix \ref{ApB}, we discuss the possibility of the appearance of a nonzero chiral isotopic density in quark matter inside neutron stars. It can be explained on the basis of the chiral separation effect in the presence of a strong magnetic field in a dense baryonic medium.

However, at the physical point ($m_0\ne 0$) the symmetry of the Lagrangian (1) under transformations from axial isotopic group $U_{AI_3}(1)$ is explicitly broken. So in the most general case with $m_0\ne 0$, $\mu_B\ne 0$ and $\mu_I\ne 0$ the initial model (\ref{1}) is invariant only under the $U_B(1)\times U_{I_3}(1)$ group. (We would like also to remark that Lagrangian (1) is invariant with respect to the electromagnetic $U_Q(1)$ group,
$U_Q(1):~q\to\exp (\mathrm{i}Q\alpha) q$,
at arbitrary values of $m_0$, where $Q={\rm diag}(2/3,-1/3)$.)

The ground state (the state of thermodynamic equilibrium) of quark matter with $n_B\ne 0$ and $n_I\ne 0$, where $n_B\equiv\vev{\hat n_B}$, $n_I\equiv\vev{\hat n_I}$,\footnote{The notation $\vev{\hat O}$ means the ground state expectation value of the operator $\hat O$.} 
both at zero and nonzero values of $m_0$ has been investigated in the framework of the NJL model (1), 
e.g., in Refs. \cite{ek1,Ebert}. However, the fact that quark matter may have a nonzero chiral isotopic charge was ignored in those papers. Recently, this gap in researches was filled in the paper \cite{Khunjua:2017mkc}, where we have studied the properties of equilibrium quark matter at $n_B\ne 0$, $n_I\ne 0$ as well as at nonzero chiral isospin charge density $n_{I5}\equiv\vev{\hat n_{I5}}\ne 0$ in the framework of the massless (3+1)-dimensional two-flavor NJL model (temperature $T$ was taken to be zero in Ref. \cite{Khunjua:2017mkc}). In contrast to this, in the present paper we consider the properties of a more realistic quark matter, i.e. at $m_0\ne 0$ and $T\ne 0$, for which all densities $n_B$, $n_I$ and $n_{I5}$ are also nonzero. The solution of this problem can be most conveniently carried out in terms of chemical potentials $\mu_B$, $\mu_I$ and $\mu_{I5}$, which are the quantities, thermodynamically conjugated to corresponding charge densities $\hat n_B$, $\hat n_I$ and $\hat n_{I5}$ presented in Eqs. (\ref{2003}) and (\ref{2004}). Therefore, when solving this problem, one can rely on the Lagrangian of the form
\begin{eqnarray}
  \bar L&=&L+\mu_{I5}\hat n_{I5}\nonumber\\
&=&\bar q\Big [\gamma^\nu\mathrm{i}\partial_\nu -m_0
+\frac{\mu_B}{3}\gamma^0+\frac{\mu_I}2 \tau_3\gamma^0+\frac{\mu_{I5}}2 \tau_3\gamma^0\gamma^5\Big ]q+ \frac
{G}{N_c}\Big [(\bar qq)^2+(\bar q\mathrm{i}\gamma^5\vec\tau q)^2 \Big
]. \label{40}
\end{eqnarray}
(Generally speaking, in this case the chiral isospin charge is no more a conserved quantity of our system. Therefore, chiral isospin chemical potential $\mu_{I5}$ is not conjugated to a strictly conserved charge. However, denoting by $\tau$ the typical time scale in which all chirality changing processes take place, one can treat $\mu_{I5}$ as the chemical potential that describes a system in thermodynamic equilibrium with a fixed value of $n_{I5}$ on a time scale much larger than $\tau$.)

Our goal is the investigation of the ground state properties (or phase structure) of the system, described by the Lagrangian (\ref{40}), and its dependence on the chemical potentials $\mu_B$, $\mu_I$ and $\mu_{I5}$ (both at zero and nonzero temperature). It is well known that all information on the phase structure of the model is contained in its thermodynamic potential (TDP). Namely, in the behavior of its global minimum point vs. chemical potentials. Moreover, the values of charge densities $n_B\equiv\vev{\hat n_B}$, $n_I\equiv\vev{\hat n_I}$ and $n_{I5}\equiv\vev{\hat n_{I5}}$ in equilibrium quark matter can be found by differentiating the TDP in the global minimum point with respect to the corresponding chemical potentials $\mu_B$, $\mu_I$ and $\mu_{I5}$, etc. In order to find the TDP of the model, we start from a semibosonized version of the Lagrangian (\ref{40}), which contains composite bosonic fields $\sigma (x)$ and $\pi_a (x)$:
\begin{eqnarray}
{\cal L}\ds =\bar q\Big [\gamma^\rho\mathrm{i}\partial_\rho - m_0 +\mu\gamma^0
+ \nu\tau_3\gamma^0+\nu_{5}\tau_3\gamma^0\gamma^5-\sigma
-\mathrm{i}\gamma^5\pi_a\tau_a\Big ]q
 -\frac{N_c}{4G}\Big [\sigma\sigma+\pi_a\pi_a\Big ].
\label{2}
\end{eqnarray}
Here, $a=1,2,3$ and also we introduced the notations $\mu\equiv\mu_B/3$, $\nu\equiv\mu_I/2$ and $\nu_{5}\equiv\mu_{I5}/2$. From the auxiliary Lagrangian (\ref{2}) one gets the equations for the bosonic fields:
\begin{eqnarray}
\sigma(x)=-2\frac G{N_c}(\bar qq);~~~\pi_a (x)=-2\frac G{N_c}(\bar q
\mathrm{i}\gamma^5\tau_a q).
\label{200}
\end{eqnarray}
Note that the composite bosonic field $\pi_3 (x)$ can be identified with the physical $\pi^0(x)$-meson field, whereas the physical $\pi^\pm (x)$-meson fields are the following combinations of the composite fields, $\pi^\pm (x)=(\pi_1 (x)\mp i\pi_2 (x))/\sqrt{2}$.
Obviously, the semibosonized Lagrangian ${\cal L}$ is equivalent to the initial Lagrangian (\ref{40}) when using the equations (\ref{200}). Furthermore, the composite bosonic fields (\ref{200})
change under the influence of transformations from the isospin $U_{I_3}(1)$ and axial isospin $U_{AI_3}(1)$ 
groups in the following manner:
\begin{eqnarray}
U_{I_3}(1):~&&\sigma\to\sigma;~~\pi_3\to\pi_3;~~\pi_1\to\cos
(\alpha)\pi_1+\sin (\alpha)\pi_2;~~\pi_2\to\cos (\alpha)\pi_2-\sin
(\alpha)\pi_1,\nonumber\\
U_{AI_3}(1):~&&\pi_1\to\pi_1;~~\pi_2\to\pi_2;~~\sigma\to\cos
(\alpha)\sigma+\sin (\alpha)\pi_3;~~\pi_3\to\cos
(\alpha)\pi_3-\sin (\alpha)\sigma.
\label{201}
\end{eqnarray}

\subsection{Thermodynamical potential. Zero temperature case.}

Starting from the auxiliary Lagrangian (\ref{2}), one obtains in the leading order of the large-$N_c$ expansion (i.e. in the one-fermion loop
approximation) the following path integral expression for the
effective action ${\cal S}_{\rm {eff}}(\sigma,\pi_a)$ of the bosonic
$\sigma (x)$ and $\pi_a (x)$ fields:
$$
\exp(\mathrm{i}{\cal S}_{\rm {eff}}(\sigma,\pi_a))=
  N'\int[d\bar q][dq]\exp\Bigl(\mathrm{i}\int{\cal L}\,d^4 x\Bigr),
$$
where
\begin{equation}
{\cal S}_{\rm {eff}}
(\sigma(x),\pi_a(x))
=-N_c\int d^4x\left [\frac{\sigma^2+\pi^2_a}{4G}
\right ]+\tilde {\cal S}_{\rm {eff}},
\label{3}
\end{equation}
The quark contribution to the effective action, i.e. the term
$\tilde {\cal S}_{\rm {eff}}$ in (\ref{3}), is given by:
\begin{eqnarray}
\exp(\mathrm{i}\tilde {\cal S}_{\rm {eff}})&=&N'\int [d\bar
q][dq]\exp\Bigl(\mathrm{i}\int\Big\{\bar q\big
[\gamma^\rho\mathrm{i}\partial_\rho -m_0+\mu\gamma^0+
\nu\tau_3\gamma^0+\nu_5\tau_3\gamma^0\gamma^5-\sigma -\mathrm{i}\gamma^5\pi_a\tau_a\big
]q\Big\}d^4 x\Bigr)\nonumber\\
&=&[\Det D]^{N_c},
 \label{4}
\end{eqnarray}
where $N'$ is a normalization constant. Moreover, in (\ref{4}) we have introduced the notation $D$,
\begin{equation}
D\equiv\gamma^\nu\mathrm{i}\partial_\nu -m_0+\mu\gamma^0
+ \nu\tau_3\gamma^0+\nu_{5}\tau_3\gamma^0\gamma^5-\sigma (x) -\mathrm{i}\gamma^5\pi_a(x)\tau_a,
\label{5}
\end{equation}
for the Dirac operator, which acts in the flavor-, spinor- as well as coordinate spaces only. Using the general formula $\Det D=\exp {\rm Tr}\ln D$, one obtains for the effective action (\ref{3}) the following expression
\begin{equation}
{\cal S}_{\rm {eff}}(\sigma(x),\pi_a(x))
=-N_c\int
d^4x\left[\frac{\sigma^2(x)+\pi^2_a(x)}{4G}\right]-\mathrm{i}N_c{\rm
Tr}_{sfx}\ln D,
\label{6}
\end{equation}
where the Tr-operation stands for the trace in spinor- ($s$), flavor-
($f$) as well as four-dimensional coordinate- ($x$) spaces, respectively.

The ground state expectation values $\vev{\sigma(x)}$ and
$\vev{\pi_a(x)}$ of the composite bosonic fields are determined by
the saddle point equations,
\begin{eqnarray}
\frac{\delta {\cal S}_{\rm {eff}}}{\delta\sigma (x)}=0,~~~~~
\frac{\delta {\cal S}_{\rm {eff}}}{\delta\pi_a (x)}=0,
\label{05}
\end{eqnarray}
where $a=1,2,3$. Just the knowledge of $\vev{\sigma(x)}$ and
$\vev{\pi_a(x)}$ and, especially, of their behaviour vs. chemical potentials supplies us with a phase structure of the model.
In the present paper we suppose that in the ground state of the system the quantities $\vev{\sigma(x)}$ and $\vev{\pi_a(x)}$ do not depend on spacetime coordinates $x$,
\begin{eqnarray}
\vev{\sigma(x)}\equiv \sigma,~~~\vev{\pi_a(x)}\equiv \pi_a, \label{8}
\end{eqnarray}
where $\sigma$ and $\pi_a$ ($a=1,2,3$) are already spatially independent constant quantities. In fact, they are coordinates of the global minimum point of the thermodynamic potential (TDP) $\Omega (\sigma,\pi_a)$.
In the leading order of the large-$N_c$ expansion 
it is defined by the following expression:
\begin{equation}
\int d^4x \Omega (\sigma,\pi_a)=-\frac{1}{N_c}{\cal S}_{\rm
{eff}}\big (\sigma(x),\pi_a (x)\big )\Big|_{\sigma
(x)=\sigma,\pi_a(x)=\pi_a} .\label{08}
\end{equation}
In what follows we are going to investigate the $\mu,\nu,\nu_{5}$-dependence of the global minimum point of the function $\Omega (\sigma,\pi_a)$ vs $\sigma,\pi_a$. Let us note that in the chiral limit (due to a $U_{I_3}(1)\times U_{AI_3}(1)$ invariance of the model) the TDP (\ref{08}) depends effectively only on the combinations $\sigma^2+\pi_3^2$ and $\pi_1^2+\pi_2^2$. Whereas at the physical point (i.e. at $m_0\ne 0$) it depends effectively on the combination $\pi_1^2+\pi_2^2$ as well as on $\sigma$ and $\pi_3$. Since in this case the relations $\vev{\sigma(x)}\ne 0$ and $\vev{\pi_3(x)} = 0$ are always satisfied (see, e.g., in Ref. \cite{Fedotov}), at $m_0\ne 0$ one can put without loss of generality $\pi_2=\pi_3=0$ in Eq. (\ref{08}), and study the TDP as a function of only two variables. For simplicity, we introduce the following $M\equiv\sigma+m_0$ and $\Delta\equiv\pi_1$ notations, and throughout the paper use the ansatz
\begin{eqnarray}
\vev{\sigma(x)}=M-m_0,~~~\vev{\pi_1(x)}=\Delta,~~~\vev{\pi_2(x)}=0,~~~ \vev{\pi_3(x)}=0. \label{06}
\end{eqnarray}
If in the global minimum point of the TDP we have $\Delta\ne 0$, then isospin $U_{I_3}(1)$ symmetry of the model is spontaneously broken down. Moreover, since at $m_0\ne 0$ chiral symmetry is explicitly broken down in the model, the $M$ coordinate of the global minimum is always a nonzero quantity. Note also that $M$ is a dynamical or constituent quark mass. In terms of $M$ and $\Delta$ the TDP (\ref{08}) reads
\begin{eqnarray}
\Omega (M,\Delta)~
&&=\frac{(M-m_0)^2+\Delta^2}{4G}+\mathrm{i}\frac{{\rm
Tr}_{sfx}\ln D}{\int d^4x}\nonumber\\
&&=\frac{(M-m_0)^2+\Delta^2}{4G}+\mathrm{i}\int\frac{d^4p}{(2\pi)^4}\ln\Det\overline{D}(p),
\label{07}
\end{eqnarray}
where
\begin{equation}
\overline{D}(p)=\not\!p +\mu\gamma^0
+ \nu\tau_3\gamma^0+ \nu_{5}\tau_3\gamma^0\gamma^5-M
-\mathrm{i}\gamma^5\Delta\tau_1\equiv\left
(\begin{array}{cc}
A~, & U\\
V~, & B
\end{array}\right )
\label{500}
\end{equation}
is the momentum space representation of the Dirac operator $D$ (\ref{5}) under the constraint (\ref{06}). The quantities $A,B,U,V$ in Eq. (\ref{500}) are really the following 4$\times$4 matrices,
\begin{equation}
A=\not\!p +\mu\gamma^0
+ \nu\gamma^0+ \nu_{5}\gamma^0\gamma^5-M;~~B=\not\!p +\mu\gamma^0
- \nu\gamma^0- \nu_{5}\gamma^0\gamma^5-M;~~U=V=-\mathrm{i}\gamma^5\Delta,
\label{80}
\end{equation}
so the quantity $\overline{D}(p)$ from Eq. (\ref{500}) is indeed a 8$\times$8 matrix whose determinant appears in the expression (\ref{07}). Based on the following general relations
\begin{eqnarray}
\Det\overline{D}(p)\equiv\det\left
(\begin{array}{cc}
A~, & U\\
V~, & B
\end{array}\right )=\det [-VU+VAV^{-1}B]=\det
[BA-BUB^{-1}V]
\label{9}
\end{eqnarray}
and using any program of analytical calculations, one can find from Eqs. (\ref{80}) and (\ref{9})
\begin{eqnarray}
\Det\overline{D}(p)=\big (\eta^4-2a\eta^2-b\eta+c\big )\big (\eta^4-2a\eta^2+b\eta+c\big )\equiv P_-(p_0)P_+(p_0),
\label{91}
\end{eqnarray}
where $\eta=p_0+\mu$, $|\vec p|=\sqrt{p_1^2+p_2^2+p_3^2}$ and
\begin{eqnarray}
a&&=M^2+\Delta^2+|\vec p|^2+\nu^2+\nu_{5}^2;~~b=8|\vec p|\nu\nu_{5};\nonumber\\
c&&=a^2-4|\vec p|^2(\nu^2+\nu_5^2)-4M^2\nu^2-4\Delta^2\nu_5^2-4\nu^2\nu_5^2.
\label{10}
\end{eqnarray}
It is evident from Eq. (\ref{10}) that the TDP (\ref{07}) is an even function over the variable $\Delta$, and parameters $\nu$ and $\nu_5$. In addition, it is invariant under the transformation $\mu\to-\mu$. \footnote{Indeed, if simultaneously with $\mu\to-\mu$ we perform in the integral (\ref{07}) the $p_0\to-p_0$ change of variables, then one can easily see that the expression (\ref{07}) remains intact. }
Hence, without loss of generality we can consider in the following only $\mu\ge 0$, $\nu\ge 0$, $\nu_5\ge 0$, and $\Delta\ge 0$ values of these quantities. Moreover in the chiral limit, the TDP (\ref{07}) is invariant with respect to the so-called duality transformation:
\begin{eqnarray}
{\cal D}:~~~~M\longleftrightarrow \Delta,~~\nu\longleftrightarrow\nu_5.
 \label{16}
\end{eqnarray}
(It is interesting to note that the dual symmetry (\ref{16}) is also an inherent property of the TDP of the model (\ref{40}) in the chiral limit and at $N_c\to\infty$, but in the (1+1)-dimensional spacetime \cite{2dim}.)  
One can find roots of the polynomials (\ref{91}) analytically, the procedure is relegated to Appendix \ref{ApA}. Four roots of $P_{+}(\eta)$ have the following form
\begin{eqnarray}
\eta_{1}=\frac{1}{2} \left(-\sqrt{r^2-4 q}-r\right),&&
~~~\eta_{2}=\frac{1}{2} \left(\sqrt{r^2-4 q}-r\right),\nonumber\\
\eta_{3}=\frac{1}{2} \left(r-\sqrt{r^2-4 s}\right),&&
~~~\eta_{4}=\frac{1}{2} \left(r+\sqrt{r^2-4 s}\right). \label{01}
\end{eqnarray}
The roots of $P_{-}(\eta)$ can be obtained by changing $b\to-b$ (changing $b\to-b$ is equivalent to $q\leftrightarrow s$),
\begin{eqnarray}
\eta_{5}=\frac{1}{2} \left(-\sqrt{r^2-4 s}-r\right)=-\eta_{4},&&~~~
\eta_{6}=\frac{1}{2} \left(\sqrt{r^2-4 s}-r\right)=-\eta_{3},\nonumber\\
\eta_{7}=\frac{1}{2} \left(r-\sqrt{r^2-4 q}\right)=-\eta_{2},&&~~~
\eta_{8}=\frac{1}{2} \left(r+\sqrt{r^2-4 q}\right)=-\eta_{1}.\label{02}
\end{eqnarray}
where $q=\frac{1}{2} \left(-2 a+r^2-\frac{b}{r}\right)
,\,\,\,s=\frac{1}{2} \left(-2 a+r^2+\frac{b}{r}\right)$, and $r$ has quite complicated form, but could be always chosen as a real one (all the details can be found in Appendix \ref{ApA}).
As a result, we have from Eq. (\ref{91}) that
\begin{eqnarray}
\Det\overline{D}(p)=\Pi_{i=1}^8(\eta-\eta_i),
\label{091}
\end{eqnarray}
where each root $\eta_i$ is invariant with respect to the duality transformation (\ref{16}). So, it is evident from Eqs. (\ref{07}) and (\ref{91}) that for the TDP one can obtain the following expression
\begin{eqnarray}
\Omega (M,\Delta)~
=\frac{(M-m_0)^2+\Delta^2}{4G}+\mathrm{i}\sum_{i=1}^{8}\int\frac{d^4p}{(2\pi)^4}\ln(p_{0}+\mu-\eta_{i}).\label{070}
\end{eqnarray}
Then, taking in account a general formula
\begin{eqnarray}
\int_{-\infty}^\infty dp_0\ln\big
(p_0-K)=\mathrm{i}\pi|K|,\label{int}
\end{eqnarray}
and using the fact that each root $\eta_i$ of Eqs. (\ref{01}) and (\ref{02}) has a counterpart with opposite sign as well as the relation $|\mu-\eta_{i}|+|\mu+\eta_{i}|=2|\eta_{i}|+2\theta(\mu-|\eta_{i}|)(\mu-|\eta_{i}|) $, one gets
\begin{eqnarray}
\Omega (M,\Delta)
&=&\frac{(M-m_0)^2+\Delta^2}{4G}-\sum_{i=1}^{4}\int\frac{d^3p}{(2\pi)^3}\big (|\eta_{i}|+\theta(\mu-|\eta_{i}|)(\mu-|\eta_{i}|)\big )\nonumber\\
&=&\frac{(M-m_0)^2+\Delta^2}{4G}-\frac{1}{2\pi^2}\sum_{i=1}^{4}\int_{0}^{\Lambda}p^2\big (|\eta_{i}|+\theta(\mu-|\eta_{i}|)(\mu-|\eta_{i}|)\big )dp.\label{26}
\end{eqnarray}
To obtain the second line of Eq. (\ref{26}), where $p\equiv|\vec p|$ and $\Lambda$ is a three-momentum cutoff parameter, we have integrated in the first line of it over angle variables. If we are interested in knowing the phase structure of the model at zero temperature, we should study just the TDP (\ref{26}) vs $M$ and $\Delta$ on the global minimum point (GMP). It is clear that at $m_0\ne 0$ the GMP of the TDP has the form $(M_0,\Delta_0)$, where $M_0$ is always a nonzero quantity. If in this case $\Delta_0\ne 0$, then we are in the charged PC phase with spontaneous breaking of the isospin $U_{I_3}(1)$ symmetry.

\subsection{Thermodynamical potential. Non-zero temperature case.}
\label{T}

Though, the effect of non-zero temperatures is quite predictable (one can expect that the temperatures just restore all the broken symmetries of the model), here we include nonzero temperatures into consideration because it is important in a number of applications. In heavy ion collisions and early Universe the temperatures are huge and its account looks inevitable, but it even makes sense in other not so apparent situations. We know that compact stars are cold and one can consider their temperatures as zero. But probably there could be scenarios in which the temperatures could be important even in the context of compact stars. For example,  their temperatures right after they are born in a supernova explosion can be as high as $T\approx10$ MeV. So it is instructive to know how robust the charged PC phase under temperature.

To introduce finite temperature into consideration, it is very convenient to use the zero temperature expression (\ref{070}) for the TDP. Then, to find the temperature dependent TDP $\Omega_T(M,\Delta)$ one should replace in Eq. (\ref{070}) the integration over $p_0$ in favor of the summation over Matsubara frequencies $\omega_n$ by the rule
\begin{eqnarray}
\int_{-\infty}^{\infty}\frac{dp_0}{2\pi}\big (\cdots\big )\to
iT\sum_{n=-\infty}^{\infty}\big (\cdots\big ),~~~~p_{0}\to
p_{0n}\equiv i\omega_n \equiv i\pi T(2n+1),~~~n=0,\pm 1, \pm 2,...,
\label{190}
\end{eqnarray}
In the expression obtained, it is possible to sum over Matsubara frequencies using the general formula (the corresponding technique is presented, e.g., in \cite{jacobs})
\begin{eqnarray}
&&\sum^{\infty}_{n=-\infty}\ln (i\omega_n-a)
=\ln\left [\exp (\beta |a|/2)+\exp (-\beta |a|/2)\right ]
=\frac{\beta
|a|}{2}+\ln\left [1+\exp (-\beta |a|)\right ],\label{C4}
\end{eqnarray}
where $\beta=1/T$. As a result, one can obtain the following expression for the TDP $\Omega_T(M,\Delta)$
\begin{eqnarray}\label{TDPT}
\Omega_T (M,\Delta)
&=&\Omega (M,\Delta)
-T\sum_{i=1}^{4}\int_{0}^{\Lambda}\frac{p^2dp}{2\pi^2}\Big\{\ln(1+e^{-\frac{1}{T}(|\eta_{i}-\mu|)})+\ln(1+e^{-\frac{1}{T}(|\eta_{i}+\mu|)})\Big\},\label{260}
\end{eqnarray}
where $\Omega (M,\Delta)$ is the TDP (\ref{26}) of the system at zero temperature. Since each root $\eta_i$ in Eq. (\ref{260}) is a dually ${\cal D}$ invariant quantity (see in Eq. (\ref{16})), it is clear that in the chiral limit the temperature dependent TDP (\ref{260}) is also symmetric with respect to the duality transformation ${\cal D}$.

Finally, it is necessary to note
that in the framework of the NJL$_4$ model the leading order of the large-$N_c$ limit is identical to the mean-field approximation. This suggests that at $T=0$, where 
fluctuations are expected to be suppressed, the results are likely to be a good approximation to the $N_c=3$ case. To be sure that this fact is also valid at nonzero temperature, one can remember, e.g., Refs. \cite{Radzhabov}, where it was shown that mean-field approximation is a rather good approximation both at zero and finite 
temperature. So not only in the limit $N_c\to\infty$ but also at finite $N_c$ thermal fluctuations are 
 not that large and probably cannot destroy the results, obtained in this approximation. Hence, in the following we may compare 
our NJL$_4$ results with lattice $N_c=3$ QCD results at $T>0$. 
Despite all this arguments one should be very cautious comparing the results obtained in different approaches with different setups and approximations, and it should be mentioned that the agreement can be qualitative and not very precise.

\subsection{Technical details}

Technically, to define the ground state of the system one should find the coordinates $(M_0,\Delta_0)$ of the global minimum point (GMP) of the TDP (\ref{26}). Since the NJL model is a non-renormalizable theory we have to use fitting parameters for the quantitative investigation of the system. We use the following, widely used parameters:
\begin{eqnarray}
m_0 = 5,5 \,{\rm MeV};\qquad G=15.03\, {\rm GeV}^{-2};\qquad \Lambda=0.65\, {\rm GeV}.\label{fit}
\end{eqnarray}
In this case at $\mu=\nu=\nu_5=0$ one gets for constituent quark mass the value $M=309\,{\rm MeV}$. Moreover, we suppose that quark chemical potentials are varied in the region $\mu<\Lambda$, $\nu<\Lambda$ and $\nu_{5}<\Lambda$. At higher values of $\mu$, $\nu$ and $\nu_{I5}$ the NJL model (\ref{40}) no longer describes a phase structure of real quark matter. The reason is rather obvious and in addition in this case it is necessary to take into account the condensation of $\rho$ mesons, color superconductivity phenomenon, etc.
Actually, even though it is shown in \cite{Cao:2015xja} that if one includes $\mu_{5}$ into consideration the transition to the color superconducting phase shifts to higher values of $\mu$, it does not in any way forbid this phenomenon and the color superconducting phase can appear in the region under consideration. So we should admit that it is interesting to include the possibility of color superconductivity but for simplicity here we will neglect it.

As our main goal of the present paper is to prove the possibility of the charged PC phenomenon in hot dense quark matter with chiral imbalance, i.e. in the framework of the NJL model (\ref{40}), the consideration of the physical quantity $n_{q}$, called quark number density, is now in order. This quantity is a very important characteristic of the ground state, especially in dynamical 
phenomena such as superfluidity. It is related to the baryon number density as $n_{q}=3n_B$ because $\mu=\mu_B/3$. In the general case this quantity is defined by the relation
\begin{eqnarray}
n_q=-\frac{\partial\Omega(M_0,\Delta_0)}{\partial\mu}, \label{37}
\end{eqnarray}
where $M_0$ and $\Delta_0$ are coordinates of the GMP of a thermodynamic potential. In addition, one can find also the density $n_I$ of isospin, $n_I=-\partial\Omega(M_0,\Delta_0)/\partial\mu_I$, as well as the chiral isospin density $n_{I5}$, $n_{I5}=-\partial\Omega(M_0,\Delta_0)/\partial\mu_{I5}$.

We distinguish the following phases that could be realized in the chirally asymmetric system under different external circumstances (the quantities $M_0$ and $\Delta_0$ below are the coordinates of the GMP of the TDP (\ref{26}) in the corresponding phase):
\begin{itemize}
\item $M_0=0; \Delta_0=0$ -- symmetrical phase. It could be realized only in the chiral limit, $m_0=0$. Usually, in this phase $ n_q\ne 0$ at $\mu\ne 0$.
\item $M_0\ne0; \Delta_0=0; n_q=0$ -- chiral symmetry breaking phase (we use for it the notation CSB). Since quark number (baryon) density is zero in this phase, sometimes it is called the ordinary baryonic vacuum.
\item $M_0\ne 0; \Delta_0=0; n_q\ne 0$ -- chiral symmetry breaking phase with nonzero quark density (below it is CSB$_{d}$ phase).
\end{itemize}
In the CSB phase the order parameter $M_0$ is usually greater than quark number chemical potential $\mu$. Moreover, $M_0$ is of order of the gap in the energy spectrum of quarks. Due to this reasons quarks cannot be created in this phase and $n_q=0$. However, with increasing of chemical potentials, it is advantageous for the system to abruptly decrease the parameter $M_0$ (see, e.g., the right panel of Fig. 5) and move into a new CSB$_d$ phase. In this case, the gap in the energy spectrum of quarks significantly decreases, which makes it possible to create quarks in the ground state. As a result, the quark number density $n_q$ is nonzero in the CSB$_d$ phase.
\begin{itemize}
\item $M_0\ne 0; \Delta_0\ne 0; n_q=0$ -- charged pion condensation phase  with zero quark density (below in all phase diagrams we use for it the notation PC) ($M_0=0$ in the chiral limit). In the charged PC phase $U_{I_3}(1)$ symmetry is spontaneously broken down. Since in this phase $n_q=0$, sometimes it is called the charged pion gas phase.
\item $M_0\ne0; \Delta_0\ne 0; n_q\ne 0$ -- charged pion condensation phase with nonzero quark density (PC$_{d}$). In the PC$_{d}$ phase $U_{I_3}(1)$ symmetry is also spontaneously broken down. \footnote{The transition between PC and PC$_d$ phases is also a first-order phase transition, as in this case the order parameter $\Delta_0$ decreases by a jump (see, e.g., the left panel of Fig. 5), and the possibility for the creation of quarks appears. Therefore, in the PC$_d$ phase the quark number density $n_q$ is nonzero. Moreover, in both phases the isospin density $n_I$ is nonzero.}
\item We use the notation ApprSYM for the approximate symmetrical phase. In the literature this phase is usually called Wigner-Weyl phase \cite{Klevansky:1992qe,Zong}. It also corresponds to a GMP of the TDP (\ref{26}), in which $M_0\ne 0$ and $\Delta_0=0$. But in contrast to the CSB and CSB$_d$ phases, dynamical quark mass $M_0$ in the ApprSYM phase drops rapidly and continuously to the current quark mass $m_0$ with increasing temperature or chemical potentials. As it follows from Eqs. (\ref{200}) and (\ref{06}), under such conditions the chiral condensate $\vev{\bar q q}$ is almost zero, and the chiral symmetry is {\it approximately} restored in the model. Moreover, at $m_0\to 0$ this phase turns into an exactly symmetrical phase with $M_0=0$. These are the reasons why we use the name ApprSYM in all phase portraits below.
\end{itemize}
Note that at zero temperature $M_0$ changes its value by a jump when there is a phase transition from different CSB or charged PC phases to the ApprSYM phase (see, e.g., in Figs 4, 5). However, at nonzero temperature there is usually a chiral crossover transition between CSB and ApprSYM phases (see in Fig. 8).

Below we present different phase portraits of the model as well as its properties in terms of this notations. \label{notations}
%


\section{Phase structure of the model}

\subsection{Exact duality in the chiral limit $(m_0=0)$ at zero temperature $(T=0)$}
\label{IIIA}

Let us first consider some equilibrium properties of the model starting from the TDP (\ref{07}) or (\ref{26}), i.e. at zero temperature, and in the chiral limit ($m_0=0$). Although this case has been investigated in details in the article \cite{Khunjua:2017mkc}, it is useful to recall the main features of the model phase structure obtained in the leading order of the large-$N_c$ expansion.

It was already noted above that in the chiral limit the TDP (\ref{07}) is invariant under the so-called duality transformation ${\cal D}$, where ${\cal D}:~M\leftrightarrow \Delta,~~\nu\leftrightarrow\nu_5$, which could be strictly seen from Eq. (\ref{10}). It means that if at some fixed values $\mu,\nu=A,\nu_5=B$ of the chemical potentials the TDP has a GMP of the form $(M=M_0,\Delta=\Delta_0)$, then at the transposed values of the isospin chemical potentials, i.e. at $\nu=B,\nu_5=A$, but at the unaltered value of $\mu$, the GMP of the TDP (\ref{07}) lies already at the point $(M=\Delta_0,\Delta=M_0)$. As a result, we see that if at $\mu,\nu=A,\nu_5=B$, e.g., the CSB phase is realized with $M=M_0\ne 0,\Delta=0$, then at the permuted (we say dually conjugated) values $\mu,\nu=B,\nu_5=A$ of chemical potentials the charged PC phase should be realized with $M=0,\Delta=M_0$, and vice versa. Hence, in the $(\nu,\nu_5)$-phase portrait all charged PC phases should be arranged mirror symmetrically to all CSB phases with respect to the line $\nu=\nu_5$. However, the symmetrical phase turns into itself under the duality transformation, and on the $(\nu,\nu_5)$-plane the line $\nu=\nu_5$ is the axis of symmetry of this phase. Just these facts are well illustrated by the $(\nu,\nu_5)$-phase diagrams of Fig. 1. There one can see three $(\nu,\nu_5)$-phase portraits of the model: the left panel corresponds to $\mu=0\,{\rm MeV}$, at the central panel $\mu=150\,{\rm MeV}$ and at the right one $\mu=200\,{\rm MeV}$. Moreover, It is clear from the phase diagrams of Fig. 1 that in dense quark matter, i.e. at $\mu>0$, $\nu_5$-chemical potential does promote the charged PC phase with nonzero quark density (there it is PC$_{\rm d}$ phase).
\begin{figure}
\includegraphics[width=1.0\textwidth]{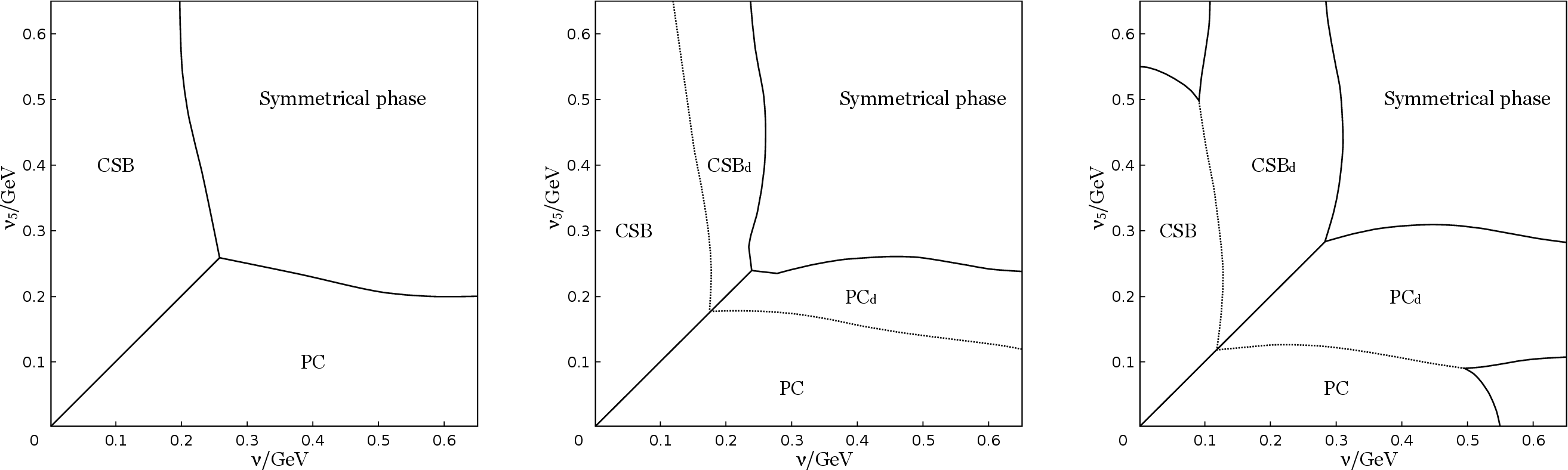}
\caption{The $(\nu,\nu_5)$-phase portraits of the model in the chiral limit $(m_0=0)$ for $\mu=0\,{\rm MeV}$ -- (a); $\mu=150\,{\rm MeV}$ -- (b); and $\mu=200\,{\rm MeV}$ -- (c).  Notations are presented in Sec. \ref{notations}.}
\label{fig1}
\end{figure}

So, in the presence of duality the knowledge of a phase of the model (\ref{40}) at some fixed values of external free model parameters $\mu,\nu,\nu_{5}$ (and at $m_0=0$) is sufficient to understand what a phase (we call it a dually conjugated) is realized at rearranged values of isospin chemical potentials, $\nu\leftrightarrow\nu_{5}$, at fixed $\mu$. Furthermore, different physical parameters such as condensates, densities, etc, which characterize both the initial phase and the dually conjugated one, are connected by the main duality transformation ${\cal D}$. For example, the chiral condensate of the initial CSB phase at some fixed $\mu,\nu,\nu_{5}$ is equal to the charged-pion condensate of the dually conjugated charged PC phase. The quark number density $n_q(\nu,\nu_{5})$ (\ref{37}) of the initial CSB phase is equal to the quark number density in the dually conjugated charged PC phase, etc.

Perhaps, the duality between CSB and charged PC phases is valid in the framework of the NJL$_4$ model under consideration only in the leading large-$N_c$ order (and at $m_0=0$). However, we think that some signs of this duality remain at the physical point of the full theory and can be observed, e.g., using lattice calculations. What gives us duality? If exact or approximate dual symmetry between different phenomena exists in the model, then, knowing the phase structure or other thermodynamic characteristics of the model in a certain region of chemical potentials, one can predict its properties in the dual-conjugated domain. For example, due to the duality between CSB and charge PC phenomena, there was no need to investigate numerically the TDP  (\ref{26}) at each point of the $(\nu,\nu_5)$-plane in order to find the phase diagrams of Fig. 1 (or the similar diagrams at other values of $\mu$). Instead, it would be sufficient to obtain a phase portrait in a more narrow region, e.g., at $\nu\ge \nu_5\ge 0$. In this case it is composed of PC, PC$_d$ and symmetrical phases (see in Fig. 1). Then one should transform each phase of it, using the mapping $\nu\longleftrightarrow\nu_5$, into a dually conjugated phase, which is already located in the region $\nu_5\ge\nu\ge 0$. At the same time we should change the name of the phase according to the rule: PC$\to$CSB, PC$_d\to$CSB$_d$ and the name of the symmetric phase under the dual transformation does not change. Thus, the duality property of the model can help to save not only the time of numerical calculations but also immediately imagine the properties of the model in previously unexplored regions of the values of chemical potentials.

There is an even more interesting use of duality. So, if we know, for example, the $(\nu,\mu)$-phase portrait of the model at fixed $\nu_5=A$, there is no need to perform detailed calculations in order to obtain its $(\nu_5,\mu)$-phase portrait at fixed $\nu=A$. To do this, it is enough to rename the $\nu$ axis of the initial phase diagram to the $\nu_5$ axis and change the name of the phases according to the rule: PC$\to$CSB, PC$_d\to$CSB$_d$ (symmetrical phase remains intact).
We call this technical procedure as the dual conjugation of a phase diagram. Hence, the $(\nu_5,\mu)$ and $(\nu,\mu)$-phase portraits are mutually conjugate to each other. However, any $(\nu,\nu_5)$-phase portrait (such as in Fig. 1) is self-dual, i.e. it is transformed into itself by the dual conjugation.

Finally note that there is another kind of duality, the duality between chiral symmetry breaking and superconductivity phenomena, which is realized in some (1+1)- and (2+1)-dimensional four-fermion theories \cite{thies,ekkz2}. But in these models the duality is a consequence of Pauli--G\"ursey symmetry of initial Lagrangians.

\subsection{Approximate duality in the case of $m_0\neq 0$ and  $T=0$}
\begin{figure}[t]
\includegraphics[width=1.0\textwidth]{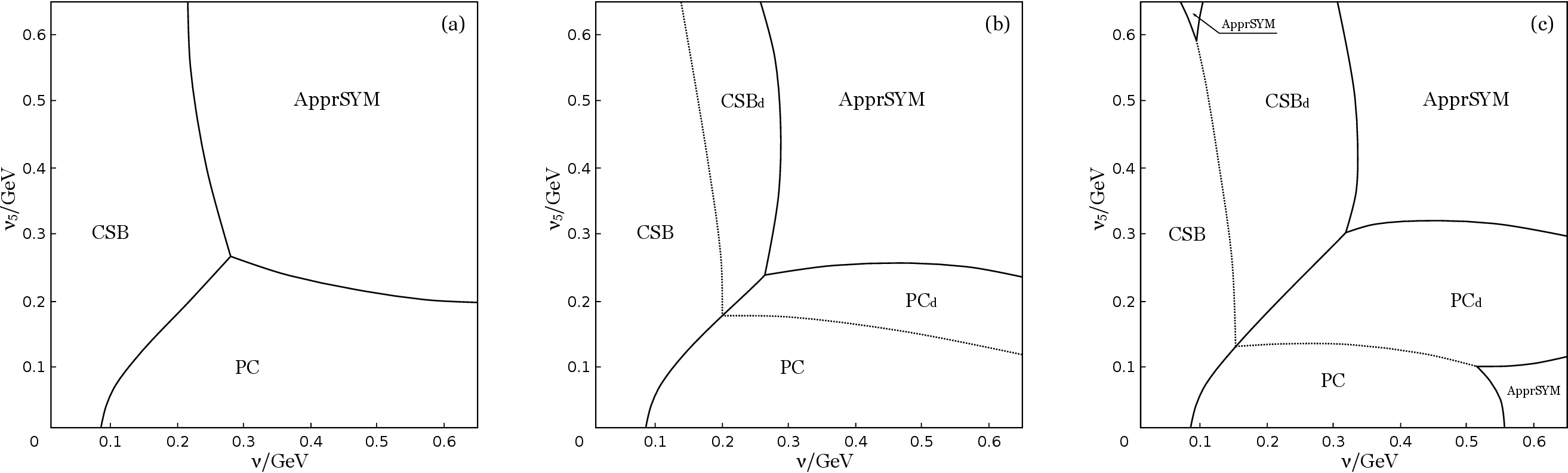}
\caption{The $(\nu,\nu_5)$-phase portraits of the model at the physical point $(m_0=5.5\, {\rm MeV})$ and at 
$\mu=0\,{\rm MeV}$ -- (a); $\mu=150\,{\rm MeV}$ -- (b); and $\mu=200\,{\rm MeV}$ -- (c). }
\end{figure}
\begin{figure}[t]
\includegraphics[width=1.0\textwidth]{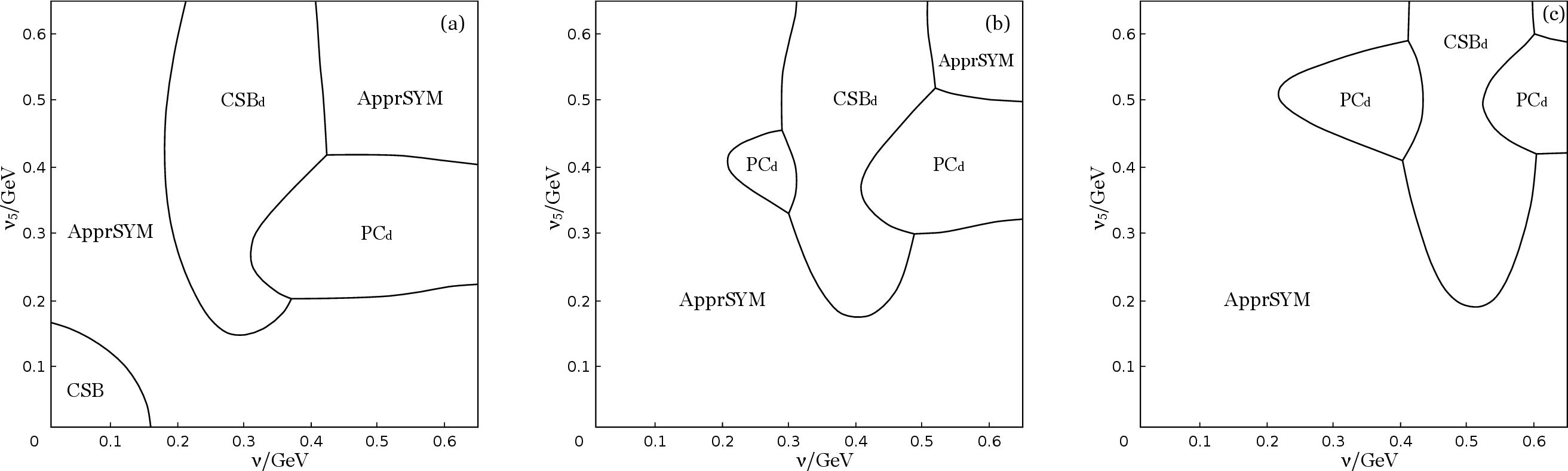}
\caption{The $(\nu,\nu_5)$-phase portraits of the model at the physical point $(m_0=5.5\, {\rm MeV})$ and at 
$\mu=300\,{\rm MeV}$ -- (a); $\mu=400\,{\rm MeV}$ -- (b); and $\mu=500\,{\rm MeV}$ -- (c). }
\end{figure}
In the present section we study the influence of a nonzero value (\ref{fit}) of the bare quark mass $m_0$ on the charged PC phase. Moreover, since at $m_0\ne 0$ the TDP (\ref{07}) is no more invariant with respect to the dual symmetry ${\cal D}$ (\ref{16}), which is exact only in the chiral limit, we will examine the question whether there are some formal signs indicating that the dual symmetry ${\cal D}$ is at least an approximate symmetry of the NJL model at $m_0\ne 0$. Among these signs are the following features of the NJL model at the physical point, when $m_0=5.5$ MeV,
\begin{enumerate}[(i)]
\item At some reliable values of the chemical potentials each $(\nu,\nu_5)$-phase portrait of the model (at some fixed $\mu$) is approximately self-dual, i.e. approximately all charged PC phases of it are arranged mirror symmetrically to all CSB phases with respect to the line $\nu=\nu_5$.
\item Each $(\nu,\nu_5)$-phase diagram has a phase (it is the ApprSYM phase), which is approximately symmetric under the transformation $\nu\leftrightarrow\nu_5$, i.e. it is arranged symmetrically with respect to the line $\nu=\nu_5$.
\item Under the dual transformation, when $\nu\leftrightarrow\nu_5$, the order parameter $M_0$ of CSB or CSB$_d$ phase is approximately equal to the order parameter $\Delta_0$ of the dually conjugated charged PC or PC$_d$ phase.
\item The quark number density $n_q$ in any phase, corresponding to the chemical potential point $(\mu,\nu=A,\nu_5=B)$, is approximately equal to quark number density $n_q$ of its dually conjugated phase that lies at the point $(\mu,\nu=B,\nu_5=A)$.
\item Each $(\nu_5,\mu)$-phase portrait (at some fixed $\nu=A$) of the model is approximately the dual ${\cal D}$ mapping of a corresponding $(\nu,\mu)$-phase portrait (at some fixed $\nu_5=A$) and vice versa.
\end{enumerate}
If these properties are inherent in the model or theory, then we say that in the model (theory) there is an approximate duality between its chiral properties and charged pion condensation phenomena.

Bearing this in mind, let us look at the $(\nu,\nu_5)$-phase portraits of Fig. 2, which are depicted for the same values of the quark number chemical potential $\mu$ as in Fig. 1. First of all note that at $\nu_5=0$ in all diagrams of Fig. 2 there is a threshold $\nu_c=m_\pi/2\approx 70$ MeV of a second order phase transition to the PC phase, which is also predicted by all known investigations \cite{Son:2000}  (including lattice calculations \cite{Gupta}). Moreover, it is easily seen from these diagrams that $\nu_5$ promotes the charged PC phase in dense quark matter (it is the phase PC$_d$ in Fig. 2 and 3) even in the case of $m_0\ne 0$. The NJL model is believed to work well at rather high baryon densities where quark matter can be realized and for low values of chemical potential $\mu<300$ MeV 
NJL model is likely to give not very trustworthy results and hadron effective model is needed. So the prediction of the generation of PC$_d$ phase at $\mu<300$ MeV is questionable and the use of hadron effective model, taking into account the presence of baryons, is needed in this region but one can consider the prediction of PC$_d$ phase generation with rather large baryon density (at $\mu>300$ MeV) to be reliable (see Fig. 3).

Concerning the above-listed duality signs (i)-(v), we see that in the region $\omega=\{(\mu,\nu,\nu_5):\mu<\Gamma (m_0),\nu<\Gamma (m_0),\nu_5<\Gamma (m_0)\}$, where $\Gamma (m_0)$ is of the order of the pion mass $m_\pi$, 
there is no sense to say about duality (even approximate), because the point (i) of this list is not fulfilled. However, as it follows from Figs. 2 and 3, outside the region $\omega$ and for all values of $\mu$, $\nu$ and $\nu_5$  restricted by the conditions $\mu<\Lambda$, $\nu<\Lambda$ and $\nu_5<\Lambda$ (the duality is even better symmetry in the region of larger values of chemical potentials but the results of NJL model in this region are not trustworthy) we see that the items (i) and (ii) are satisfied.

To have a more precise picture, let us take a look at the Figs. 4 and 5, where the Gaps $M_0,\Delta_0$ and baryon density $n_B$ vs. $\nu_5$ and $\nu$  for 
are depicted. 
It follows from these pictures that if we go from the phase, corresponding, e.g., to a chemical potential set $(\mu=200,\nu=350,\nu_5=A)$ MeV, to the dually ${\cal D}$ conjugated phase with $(\mu=200,\nu=A,\nu_5=350)$ MeV (or vice versa), then pion condensate $\Delta_0$ in the charged PC phase is approximately the same as dynamical quark mass $M_0$ in the dually conjugated CSB phase (compare the left and right panels of Fig. 5) and baryon density $n_B$ is not changed (approximately). In the dually conjugated points of the ApprSYM phase both $n_B$ and dynamic quark mass $M$ are not changed, approximately. The same conclusions one can obtain from Fig. 4 for $\mu=260$ MeV when two phases, CSB and PC, are present. Hence, the items (iii) and (iv) of the list of duality signs are also satisfied.

Finally, comparing, e.g., the $(\nu,\mu)$-phase diagram at fixed $\nu_5=200$ MeV and the $(\nu_5,\mu)$-phase diagram at fixed $\nu=200$ MeV (see in Fig. 6), we see that qualitatively they are dually ${\cal D}$ conjugated to each other at a rather low values of $\mu\lesssim 200$ MeV, i.e in this region of each diagram of Fig. 6 one can perform the following axis and phase renaming,  $\nu\leftrightarrow\nu_5$, CSB$\leftrightarrow$PC and CSB$_d\leftrightarrow$PC$_d$ (the ApprSYM phase does not change its name by the duality transformation), in order to obtain (approximately) the corresponding region of another diagram of Fig. 6. This conclusion agrees with phase portraits of Fig. 2 for moving along the lines $\nu=200$ MeV (or $\nu_5=200$ MeV) of these diagrams we intersect just the phases shown in Fig. 6 at low $\mu$. In addition, it is easy to see that there is a duality ${\cal D}$ between diagrams of Fig. 6 in the regions, where $\nu\gtrsim 200$ MeV (left panel) and $\nu_5\gtrsim 200$ MeV (right panel). So the item (v) of the list of duality signs is also satisfied.

In conclusion of this section, we can say that the duality between the phenomena of CSB and a charged PC, inherent for this model in the chiral limit at $N_c\to\infty$, is approximately fulfilled even at $m_0\ne 0$, but only for the points $(\mu,\nu,\nu_5)$ of the chemical potential space from the region, in which $\mu,\nu,\nu_5<500$ MeV and at the same time $(\mu,\nu,\nu_5)\not\in\omega$, where
$\omega=\{(\mu,\nu,\nu_5):\mu,\nu,\nu_5<\Gamma (m_0)\}$ (here $\Gamma (m_0)\sim m_\pi$
).  \label{physpoint}
\begin{figure}
\includegraphics[width=\textwidth]{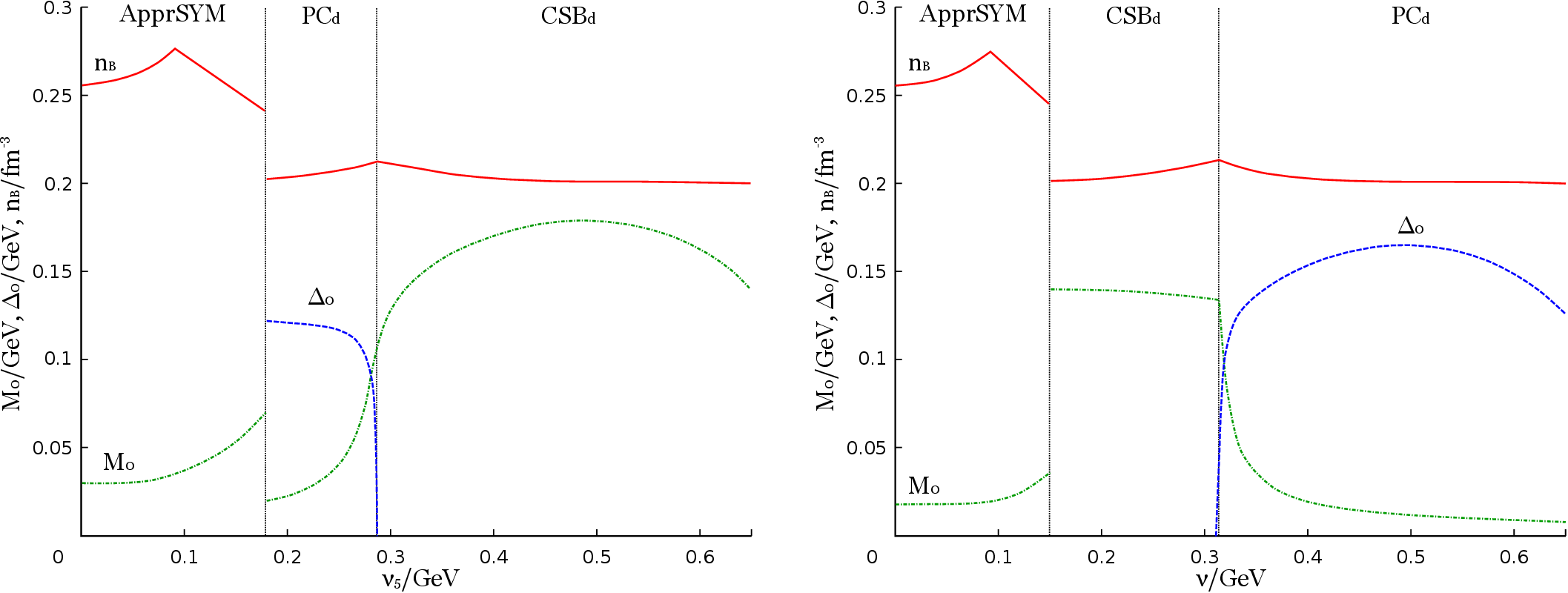}
\caption{ The Gaps $M_0$ and $\Delta_0$  vs. $\nu_5$ at $\mu=260\,{\rm MeV}$  and $\nu=300$ MeV (left panel). The same quantities $M_0$ and $\Delta_0$ vs $\nu$ at $\mu=260\,{\rm MeV}$  and $\nu_5=300$ MeV (right panel).} 
\end{figure}
\begin{figure}[t]
\includegraphics[width=0.9\textwidth]{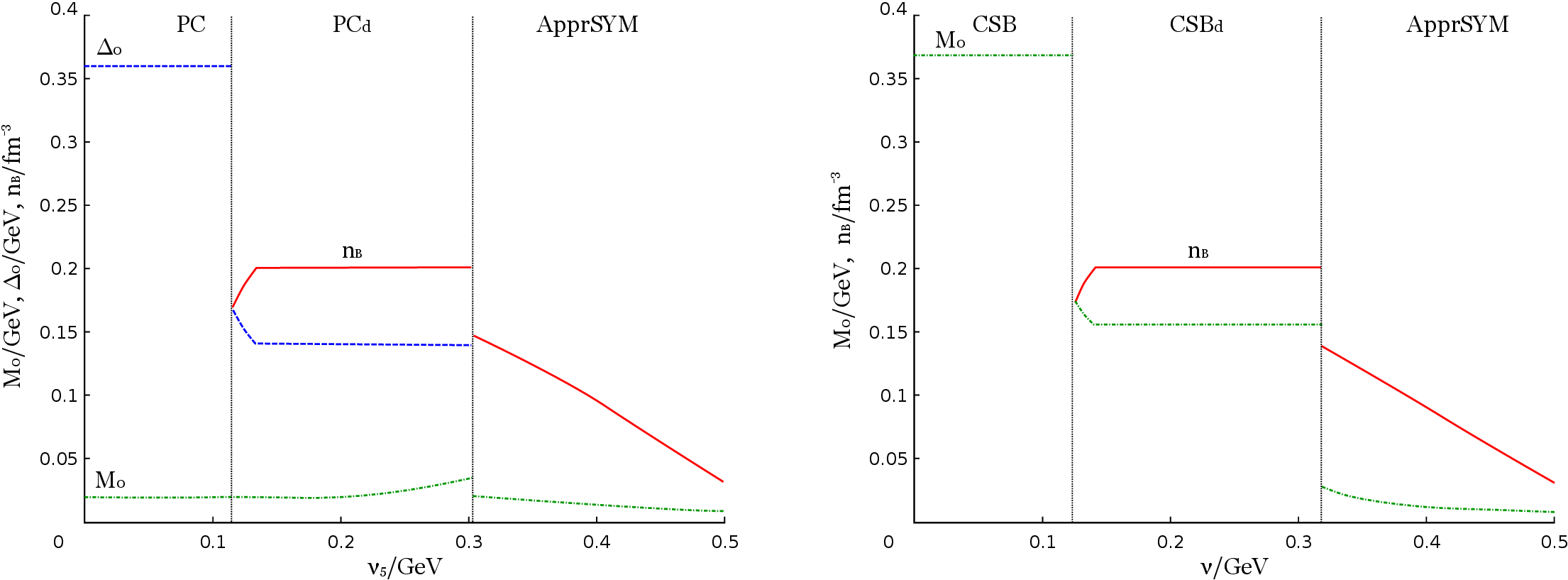}
\caption{The Gaps $M_0,\Delta_0$ and baryon density $n_B$ vs. $\nu_5$ at $\mu=200\,{\rm MeV}$  and $\nu=350$ MeV (left panel). The same quantities $M_0,\Delta_0$ and $n_B$ vs $\nu$ at $\mu=200\,{\rm MeV}$  and $\nu_5=350$ MeV (right panel).}
\label{fig4}
\end{figure}

\subsection{Phase portrait at the physical point ($m_0=5.5$ MeV) and nonzero temperature $(T\ne 0)$}
\label{IIIC}

Though, the effect of nonzero temperatures is quite predictable (indeed, one can expect that the temperature just restores all the broken symmetries of the model), we investigate nonzero temperature case because it is important in a number of applications. We know that compact stars are cold and one can consider their temperatures as zero, but probably there could be scenarios in which the temperatures could be important even in the context of compact stars. So it is instructive to know how robust the {\rm PC$\rm{_d}$-phase under the influence of temperature and chiral imbalance.

To clarify this issue, we calculated two $(\nu,T)$-phase diagrams of the model at $\mu=0$ (in order to compare our results with lattice investigations) and at different values of $\nu_5$. In Fig. 7 (left panel) one can see this diagram at $\nu_5=0$, whereas in the right panel it is at $\nu_5=200$ MeV. Note that the phase portrait at $\mu=0$ and $\nu_5=0$ is in accordance with the same phase portrait obtained within first principle lattice calculations \cite{Gupta}. Also, as one could expect, it is clear from Fig. 7 that temperature restores broken $U_{I_3}(1)$-symmetry at some rather high critical values $T_c^{PC}$, where charged PC phase is disappeared. \footnote{In the points of the boundary between ApprSYM and charged PC phases in Figs 7, 8 there is a second-order phase transition, whereas the dashed line in each of these figures represents the so-called pseudo-critical temperature, which characterizes the so-called cross-over region between CSB and ApprSYM phases.} (Of course, at fixed values of $\mu$ and $\nu_5$ the critical temperature $T_c^{PC}$ depends strongly on the isospin chemical potential $\mu_I\equiv 2\nu$.) Moreover, as it follows from Fig. 7, $T^{PC}_c$ vs. $\nu_5$ (at $\mu=0$) drops from values $T^{PC}_c\approx 200$ MeV at $\nu_5=0$ to values $T^{PC}_c\approx 100$ MeV at $\nu_5=200$ MeV (compare left and right panels of Fig. 7), i.e. when $\nu_5$ increases the region of the PC phase is shrinked in the phase portrait of the model (this fact is in accordance with the phase diagrams of Fig. 2), but nevertheless the charged PC is a quite robust effect vs temperature at $\mu=0$.

Finally, we would like to note that in addition to the list of signs (i)-(v) (see in the previous section \ref{physpoint}) indicating on the presence in the NJL model (\ref{40}) at $m_0\ne 0$ of an approximate dual symmetry between CSB and charged pion condensation at $T=0$, the similar approximate dual correspondence between phase diagrams exists also at nonzero temperature. For example, two diagrams of Fig. 8, the $(\nu,T)$-phase portrait at fixed $\mu=200$ MeV and $\nu_5=200$ MeV (left panel) and the $(\nu_5,T)$-phase portrait at fixed $\mu=200$ MeV and $\nu=200$ MeV (right panel), can be considered as a dually conjugated to each other. Indeed, applying to each of these diagrams the dual mapping, i.e. the following replacements $\nu\leftrightarrow\nu_5$, CSB$\leftrightarrow$PC and CSB$_d\leftrightarrow$PC$_d$, it is possible to obtain approximately another diagram. So dual mapping of a well-known phase portraits can be used in order to predict (approximately) a phase structure of the model at $m_0\ne 0$ in the dually conjugated region, i.e. at $\nu\leftrightarrow\nu_5$.

Last but not least conclusion from Fig. 8 (in addition to the analysis of Fig. 7) is that the charged PC is also a rather temperature stable effect in dense quark matter (at $\mu>0$).
\begin{figure}
\includegraphics[width=0.9\textwidth]{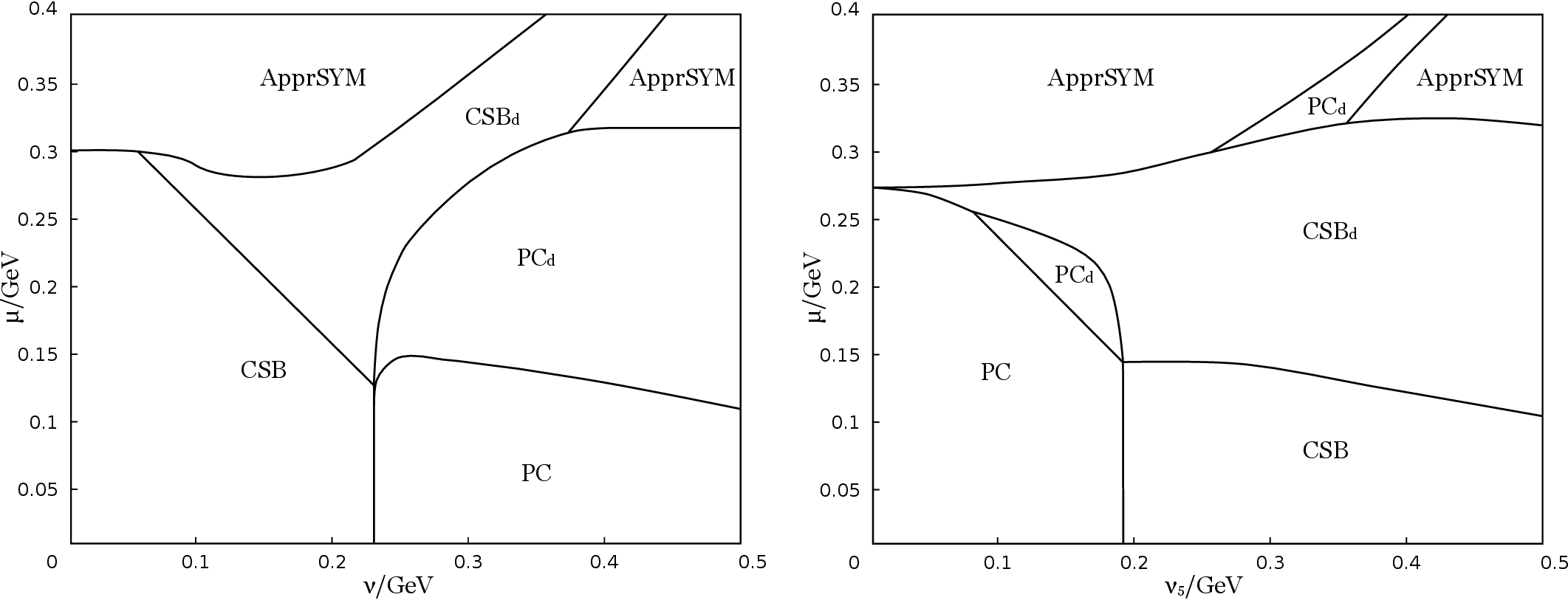}
\caption{The ($\nu,\mu$)-phase portrait at $\nu_5=200\,{\rm MeV}$ (left figure) and the ($\nu_5,\mu$)-phase portrait at $\nu=200\,{\rm MeV}$ (right figure).}
\label{fig6}
\end{figure}

\subsection{Pseudo-critical temperature $T_c(\nu_{5})$ in the NJL$_4$ model: comparison with lattice QCD and other approaches}
\begin{figure}
\includegraphics[width=0.9\textwidth]{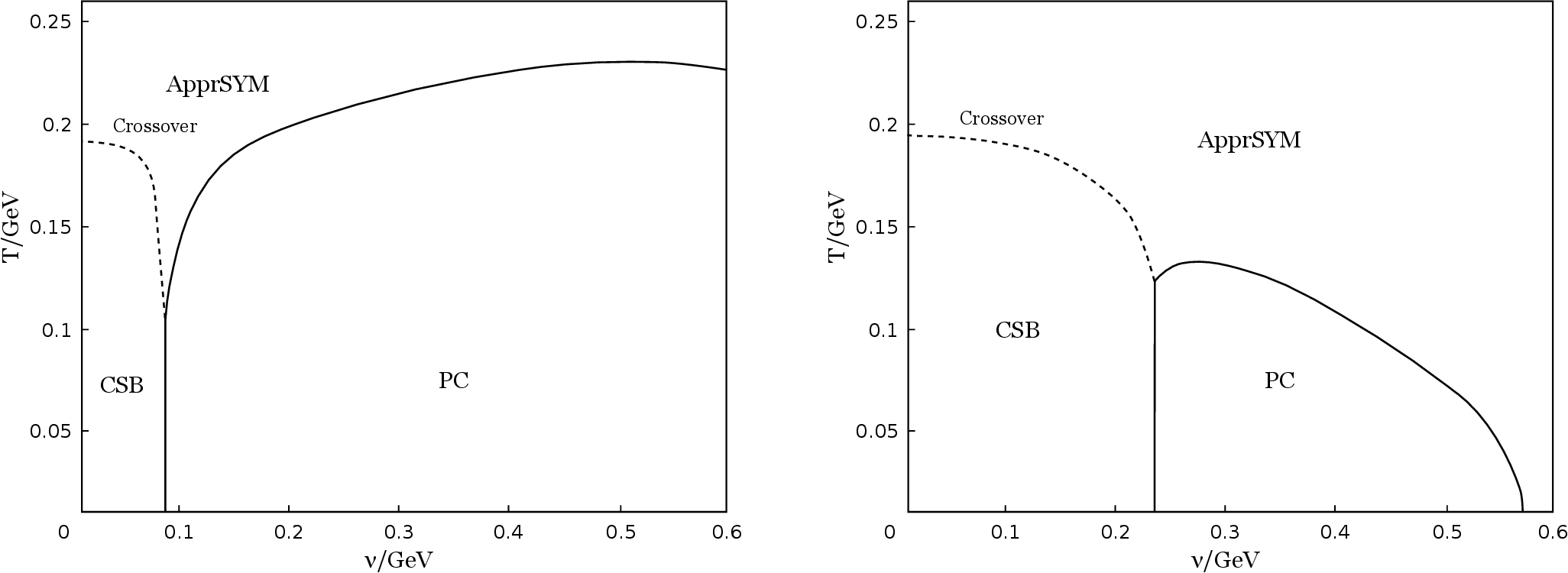}
\caption{The ($\nu,T$)-phase portraits at $\mu=\nu_5=0\,{\rm MeV}$ (left figure) and $\mu=0 \,{\rm MeV},\nu_5=200 \,{\rm MeV}$ (right figure). }
\label{fig7}
\end{figure}
\begin{figure}
\includegraphics[width=0.9\textwidth]{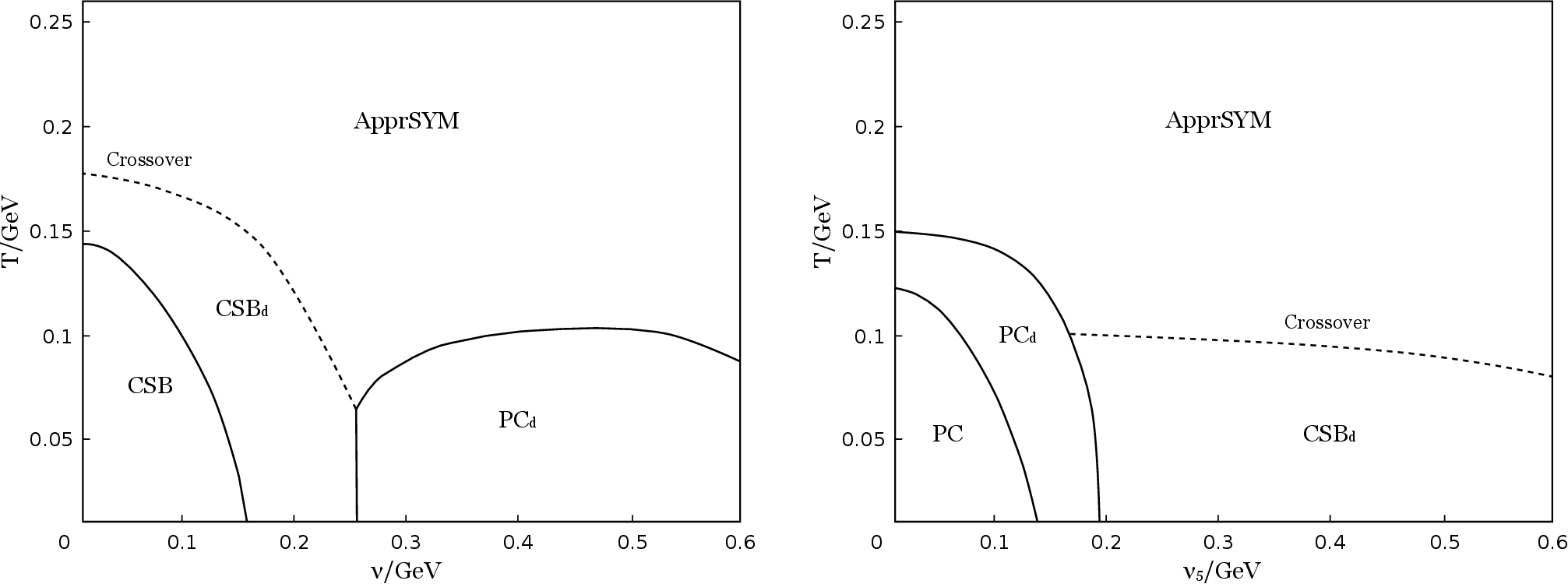}
\caption{The ($\nu,T$)-phase portraits at $\mu=200\,{\rm MeV}, \nu_5=200 \,{\rm MeV}$ (left panel) and at $\mu=200\,{\rm MeV}, \nu=200 \,{\rm MeV}$ (right panel), respectively. The dashed line corresponds to a cross-over region of the phase diagram.}
\label{fig8}
\end{figure}

A rather significant part of the previous section was devoted to the consideration of the critical temperature $T^{PC}_c$ (at $\mu=0$ and different fixed values of $\nu$) of the second-order phase transition from the charged PC to the ApprSYM phase, as well as its dependence on $\nu_5$. In addition to this, in the present section we will study in the framework of the NJL$_4$ model the behavior (at $\mu=0$ and $\nu=0$) of the pseudo-critical temperature $T_c(\nu_{5})$, which characterizes the so-called chiral cross-over region of the phase diagram (see, e.g.,  Figs. 7, 8 where this region is arranged around the dashed lines). It separates the low-temperature phase with CSB and a (partially) chirally restored ApprSYM phase, which corresponds at high temperatures to quark-gluon plasma. In the cross-over region different physical parameters, such as the dynamical quark mass $M_0$ etc, of the CSB phase smoothly (without jumps), but rather sharply go over to the corresponding parameters of the ApprSYM phase. Therefore, in this region, there occurs not a true phase transition with corresponding critical temperature, etc., but rather a {\it pseudo}-phase transition (cross-over) characterized by a {\it pseudo}-critical temperature $T_c\equiv T_c(\mu,\nu,\nu_5)$, etc. (In Figs. 7, 8 the pseudo-critical temperature $T_c$ is represented by dashed lines.) Here we study the behavior of the pseudo-critical temperature only as a function of $\nu_5$ and at fixed $\mu=0$, $\nu=0$. That is, we investigate the quantity  denoted by $T_c(\nu_{5})$,
\begin{eqnarray}
T_c(\nu_{5})\equiv  T_c(\mu,\nu,\nu_5)\Big |_{\mu=0;\nu=0}.\label{pseudo}
\end{eqnarray}
In particular, it is clear from Fig. 7 that $T_c$ at $\nu_{5}=0$ MeV (left panel of Fig. 7) is slightly smaller than $T_c$ at $\nu_{5}=200$ MeV, which can be found at the right panel of Fig. 7. The plot of the function $T_c(\nu_{5})$ vs $\nu_5$ is presented in Fig. 9. However, before comparing these our results on the pseudo-critical temperature with the predictions, obtained in the framework of other effective models and lattice QCD calculations, it is necessary to make a few remarks.

Strictly speaking, so far nobody has investigated the function  $T_c(\nu_{5})$ (\ref{pseudo}) both in the NJL model and other approaches. The matter is that in the most general case, the chiral asymmetry of dense quark matter is described by two chemical potentials, chiral $\mu_5$ and chiral isospin $\mu_{I5}\equiv 2\nu_5$ chemical potential. \footnote{In general, chiral imbalance of dense quark  matter is characterized by two densities, chiral isospin $\hat n_{I5}=\frac 12\left (\hat n_{u5}-\hat n_{d5}\right )$ and chiral density $\hat n_5=\hat n_{u5}+\hat n_{d5}$ (see Introduction for notations). Alternatively, it can be described by corresponding chemical potentials $\mu_{I5}$ and $\mu_5$, which are the quantities thermodynamically conjugated to $\hat n_{I5}$ and $\hat n_{5}$, respectively.} The first, $\mu_5$, is usually used when isotopic asymmetry of quark matter is absent, i.e. in the case $\mu_I=0$ \cite{Ruggieri}. The second, $\mu_{I5}$, might be taken into account when, in addition to chiral, there is also isotopic asymmetry of matter, in which charged PC phenomenon can be observed, etc. \cite{Khunjua:2017mkc,Khunjua:2018sro}. And up to now the behavior of a pseudo-critical temperature of the cross-over region as a function of only the chiral chemical potential $\mu_5$ was investigated in different approaches at fixed $\mu=0$, $\mu_I=0$ and $\nu_5=0$ \cite{Fukushima2,Ruggieri2,Frasca}. That is the possibility of the existence of a quark system with nonzero chiral isospin imbalance was ignored in these works. (In this particular case we use for a pseudo-critical temperature of the NJL$_4$ model the notation $T_c\equiv\widetilde T_c(\mu_{5})$. Do not confuse with the expression (\ref{pseudo}), which in fact corresponds to a pseudo-critical temperature, obtained in another limiting case of an external parameter set of the NJL$_4$ model, $\mu=0$, $\nu=0$, $\mu_5=0$ and for arbitrary values of $\nu_5$.)
\begin{figure}
\includegraphics[width=0.6\textwidth]{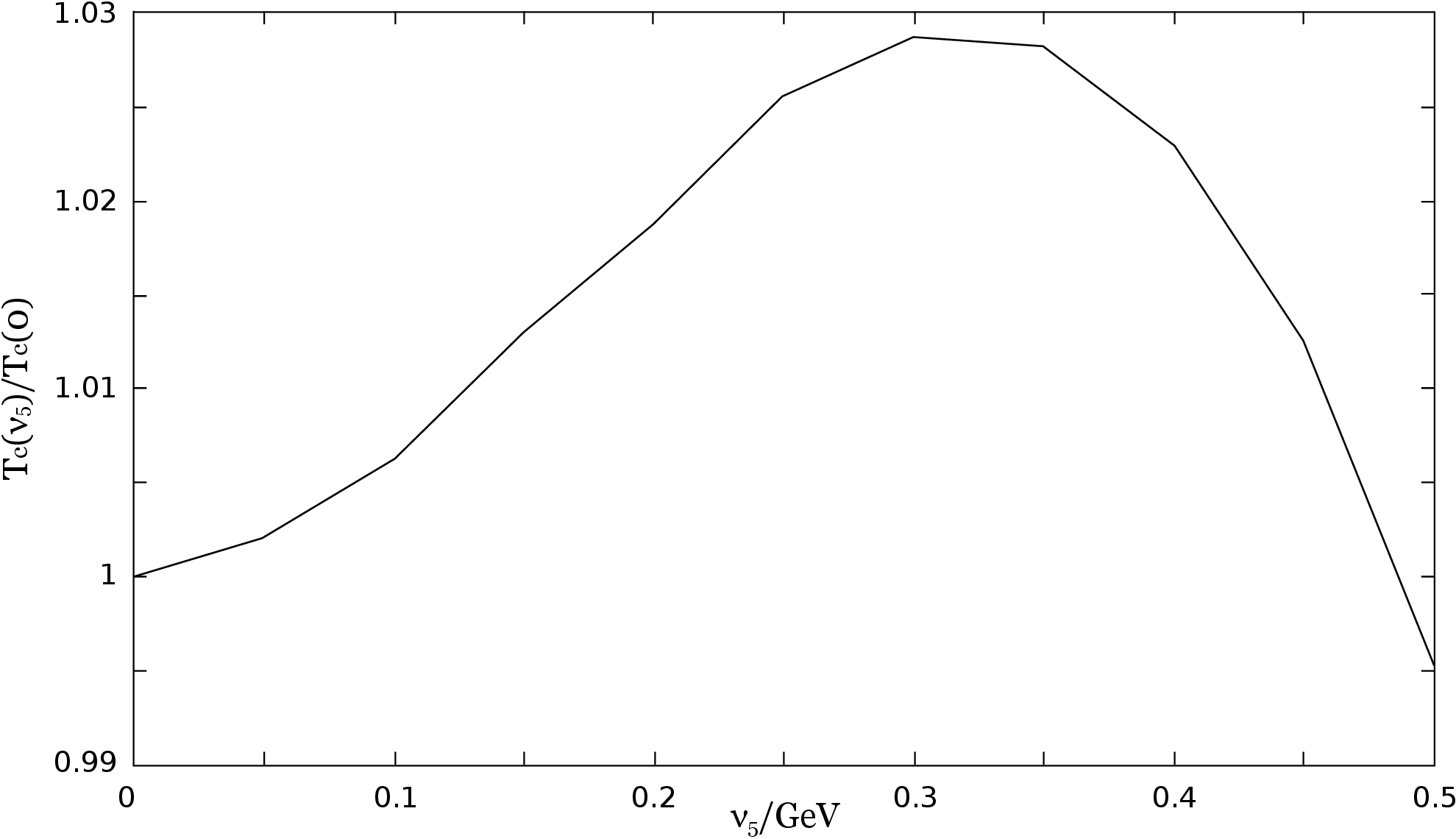}
\caption{Pseudo-critical temperature $T_c(\nu_{5})$ (\ref{pseudo}) as a function of $\nu_5$. }
\label{fig9}
\end{figure}
So in order to compare our results on the pseudo-critical temperature with other approaches to this quantity, we need formally to find in the model under consideration the behavior of a pseudo-critical temperature vs $\mu_5$, i.e. the quantity $T_c\equiv\widetilde T_c(\mu_{5})$.

Recall that for simplicity, in the present paper we study the phase structure of the NJL$_4$ model (\ref{40}) only in the case $\mu_5=0$ with other nonzero chemical potentials, however in the recent paper \cite{Khunjua:2018sro} the phase structure of this model was investigated in the chiral limit in a more general case with all four nonzero chemical potentials $\mu, \mu_5, \mu_I, \mu_{I5}$.
In particular, it was established in \cite{Khunjua:2018sro} that in addition to the dual symmetry (\ref{16}) the TDP of the NJL$_4$ model is invariant with respect to a transformation ${\cal D}_M$,
\begin{eqnarray}
{\cal D}_M:~\mu_5\longleftrightarrow\nu_5\,,~~ \Delta=0.\label{pseudo3}
\end{eqnarray}
In Ref. \cite{Khunjua:2018sro} the symmetry (\ref{pseudo3}) is called constrained duality (due to the relation $\Delta=0$). Note that, in contrast to the main duality relation ${\cal D}$ (\ref{16}), which is only an approximate symmetry between CSB and charged PC phenomena at the physical point (see in the sections above), the constrained duality ${\cal D}_M$ is an exact symmetry of phase portraits of the model even at $m_0\ne 0$. It means that in all phase diagrams of the present paper, obtained for $\mu_5=0$ and $\nu_5\ge 0$, one can treat $\nu_5$ outside of the charged PC phase as a chiral chemical potential $\mu_5$. Moreover, just due to this additional dual symmetry the following relation between pseudo-critical temperatures of the NJL$_4$ model is valid,
\begin{eqnarray}
T_c(\nu_{5})\Big |_{\nu_5=\mu_5}= \widetilde T_c(\mu_{5}), \label{pseudo2}
\end{eqnarray}
where $T_c(\nu_{5})$ and $\widetilde T_c(\mu_{5})$ are the pseudo-critical temperatures of this model in the cases $\mu=\nu=\mu_5=0$ and $\mu=\nu=\nu_5=0$, respectively. The duality can be used only in the case of $\Delta=0$ and this assumption was shown to be true in these cases in \cite{Khunjua:2018sro}. So it can be used in the considered cases, $\mu=\nu=\mu_5=0$ and $\mu=\nu=\nu_5=0$.

Just the relation (\ref{pseudo2}) gives us the possibility to compare our results with previous predictions for the (pseudo-)critical temperature $\widetilde T_c(\mu_{5})$ obtained in the framework of different effective models \cite{Ruggieri,Fukushima2,Ruggieri2,Frasca}, using the Dyson-Schwinger equations \cite{Xu,Wang} and lattice calculations \cite{Braguta:2015zta,Braguta:2015owi,Yamamoto}. It should be noted that all these studies do not provide a well-defined consistent prediction for the behavior of $\widetilde T_c(\mu_{5})$ vs $\mu_5$, and rather contradict each other. For example, the works \cite{Ruggieri,Fukushima2} predict decreasing of $\widetilde T_c(\mu_{5})$ with growing $\mu_5$, whereas in the works \cite{Xu,Frasca} there is an opposite picture.
The situation has been partially clarified in the works \cite{Yu,Ruggieri2,Farias,Cui}, where it has been shown that in the framework of the effective models, regularisation scheme could play a crucial role in the behaviour of the pseudo-critical temperature $\widetilde T_c(\mu_{5})$. 
There are several approaches to regularization of effective models. For example, one can regularize the whole TDP $\Omega_T(M,\Delta)$ (\ref{TDPT}), including both the vacuum  $\Omega(M,\Delta)$ and thermal terms of Eq. (\ref{TDPT}), or regularize only the vacuum term $\Omega(M,\Delta)$, etc. It was shown in Refs. \cite{Yu,Farias,Cui} that the behaviour of the $\widetilde T_c(\mu_{5})$ in the NJL model depends strongly on the scheme that used. And in the present paper, as it follows from Eq. (\ref{TDPT}), we regularize the whole TDP $\Omega_T(M,\Delta)$ using in Eq. (\ref{TDPT}) the so-called hard-cutoff regularization scheme when the integration region of the thermal part of the TDP is restricted by the cutoff parameter $\Lambda$.

Taking into account the relation (\ref{pseudo2}), it is easy to see from the plot of Fig. 9, that in this regularization scheme the pseudo-critical temperature $\widetilde T_c(\mu_{5})$ of the NJL model increases for $\mu_5<\mu^\ast_5 \lessapprox 400$ MeV. Above this value it drops down, but at $\mu_5>\mu^\ast_5$ the NJL$_4$ model, in our opinion, does not provide very reliable predictions, because $\mu_5$ is near the cutoff $\Lambda$. The similar quantity has been investigated in the NJL$_4$ model with the same regularization scheme but in the chiral limit \cite{Yu}. In this case $\widetilde T_c(\mu_{5})$ is no more a pseudo-critical but rather a critical temperature of a 2nd order phase transition from CSB to symmetrical phase. In contrast to Ref. \cite{Yu}, we study the NJL$_4$ model at the physical point ($m_0\ne 0$). However, the behavior of $\widetilde T_c(\mu_{5})$ vs $\mu_5$ in both cases is qualitatively the same.

The selection of such a regularization scheme is supported and justified by several things.
Namely, using the first principle lattice calculations, it was shown that the $\widetilde T_c(\mu_{5})$ increases with $\mu_5$. Then, there is a qualitative description of the mechanism leading to an increase in the pseudo-critical temperature $\widetilde T_c(\mu_{5})$. It is based on the  Fermi-sphere treatment \cite{Braguta} and backed up with the results achieved in the framework of different non-perturbative methods \cite{Xu,Wang}. And finally, and maybe most importantly in the context of the present work, we are guided by predictions for $\widetilde T_c(\mu_{5})$, which follow from the duality symmetries (\ref{16}) and (\ref{pseudo3}) of the model. Indeed, it is well established that in the $(\nu,T)$-phase portrait the critical temperature $T_c^{PC}$ at which isospin symmetry is restored increases with the chemical potential $\nu$ (see, e.g., the left panel in Fig. 7). Applying to this diagram an (approximate) duality transformation ${\cal D}$ (\ref{16}), we obtain a $(\nu_5,T)$-phase diagram corresponding to $\nu=0$, $\mu=0$ and $\mu_5=0$ with horizontal $\nu_5$ axis as well as with the CSB phase at $\nu_5\gtrapprox 0.1$ GeV. On the boundary between the ApprSYM and CSB phases there will most likely be a cross-over region  with a pseudo-critical temperature $T_c(\nu_5)$ (\ref{pseudo}) that, due to the approximate dual symmetry ${\cal D}$, should increase vs $\nu_5$, as it does $T_c^{PC}$ vs $\nu$ in Fig. 7. And finally, applying to this phase diagram the constrained duality transformation ${\cal D}_M$ (\ref{pseudo3}), we obtain a $(\mu_5,T)$-phase diagram corresponding to $\nu=0$, $\mu=0$ and $\nu_5=0$ from which it is clear that $\widetilde T_c(\mu_{5})$ also rises vs $\mu_5$. It is this qualitative analysis based on the duality properties of the NJL$_4$ model that is confirmed by Eq. (\ref{pseudo2}) along with the plot of Fig. 9. The duality is only approximate but we also saw in the previous sections that it is a good approximation for values of chemical potential larger than approximately pion mass.

Note that in lattice approach to QCD the simplest $(\nu,T)$- and $(\mu_5,T)$-phase diagrams at $\mu=0$ are well investigated. Moreover, they are in accordance with the similar phase diagrams, obtained in the framework of the NJL model (although it should be mentioned that the agreement can be not very precise because in NJL model the results were obtained in the large-$N_c$ limit and in lattice QCD with $N_c=3$). But in the last approach, as it follows from above consideration, these phase portraits are (approximately) dually conjugated to each other.  Consequently, the same connection can exist between these phase diagrams in real QCD. So there is a solid foundation, the lattice QCD simulations, which allows us to hope that duality between CSB and charged PC phenomena is one of the properties of real dense quark matter.  
\label{IIID}

\section{Summary and Discussion}

In this paper the influence of isotopic and chiral imbalance on phase structure of hot/cold dense quark matter has been
investigated at the physical point (i.e. at nonzero current quark mass $m_0$) in the framework of the (3+1)-dimensional NJL model with two quark flavors in the large-$N_{c}$ limit ($N_{c}$ is
the number of colors). Dense matter means that our consideration has been performed at nonzero baryon $\mu_B$ chemical
potential. Isotopic and chiral imbalance in the system were accounted for by introducing isospin $\mu_I$ and chiral isospin
$\mu_{I5}$ chemical potentials (see Lagrangian (\ref{40})).
Of course one knows that current quark masses of $u$ and $d$ quarks (the ones that we considered in the paper) are rather small and, in general, the chiral limit is a very good approximation. But sometimes although small but nonzero masses can change some aspects of the phase diagram. For example, charged pion condensation (PC) phase in the chiral limit and at $T=0$ starts from infinitesimally small values of isospin chemical potential $\mu_I$, but when one takes into account quark masses, then it shifts the charged PC to the values of $\mu_I$ larger than pion mass $m_\pi\approx 140$ MeV (compare diagrams of Figs 1 and 2).
The phase structure of cold dense quark matter in the chiral limit has been obtained in \cite{Khunjua:2017mkc,Khunjua:2018sro}, where it has been shown that chiral isospin chemical potential $\mu_{I5}$ generates charged PC in dense quark matter and there is a duality correspondence between CSB and charged PC phenomena in the leading order of the large-$N_{c}$ approximation. The goal of the present paper is the extension of this consideration to a more physical case of the NJL$_4$ model with nonzero current quark masses. This allows us to draw more accurate phase diagram and perform comparison with lattice QCD. Moreover, we take into account finite temperatures, which give us a chance to consider the results in the context of heavy ion collision experiments in which  temperatures are always rather large. Even in the context of neutron stars it can be interesting to consider the case of finite temperature. It has been found that the duality between CSB and charged PC phenomena
observed in \cite{Khunjua:2017mkc, Khunjua:2018sro} in the chiral limit (where it was exact) is valid with good accuracy even in the physical point. It has been also shown that temperature does not spoil the duality as well.

We have studied the full ($\mu_B$, $\mu_I$, $\mu_{I5}$, $T$)-phase diagram of quark matter in terms of the NJL$_{4}$ model with $m_0\ne 0$. This general consideration is not feasible in the lattice QCD simulations, mainly due to the famous sign problem which does not allow for the consideration of finite baryon densities (nonzero baryon chemical potential $\mu_B$).  But contrary to the case of non-zero baryon chemical potential, simulations with non-vanishing isospin $\mu_I$ and chiral $\mu_{5}$ chemical potentials are not hampered by a sign problem and some particular cases have been considered on the lattice. For example, the ($\mu_I$, $T$)-phase diagram at zero values of $\mu_B, \mu_{I5}, \mu_5$ chemical potentials is well established as in lattice QCD as well as in different effective models and a rather good agreement can be observed between this different approaches. And there are lattice QCD simulations of the quark matter with only nonzero chiral chemical potential $\mu_{5}$ in terms of as  $SU_c(2)$ QCD (two-colour QCD) \cite{Braguta:2015zta} as well as real $SU_c(3)$ QCD (three-colour QCD) \cite{Braguta:2015owi},
where the catalysis of chiral symmetry breaking by chiral chemical potential has been established. Namely, it has been shown that chiral condensate and (pseudo)critical temperature (the temperature at which the chiral condensate drops) grows with increase of chiral $\mu_{5}$.
In this paper, as well as in Ref. \cite{Khunjua:2018sro}, we have supported these conclusions by effective NJL model considerations. In paricular, the plot of the pseudo-critical temperature $\widetilde T_c(\mu_{5})$ vs chiral chemical potential (see Fig. 9 and take into account Eq. (\ref{pseudo2}))  has been drawn and it was shown that this quantity rises with the raise of  $\mu_{5}$ and the behaviour is rather similar to the results of the lattice QCD simulations.

So let us gaze at all this from the general picture viewpoint. We have two lattice simulation results,  ($\mu_I$, $T$)- and  ($\mu_5$, $T$)-phase diagrams. These phase diagrams have been also obtained in the NJL model and the results are in a good agreement with lattice QCD simulations. But in terms of NJL model we can consider the general case and we know that there is  the duality ${\cal D}$ (\ref{16}) (it is exact in the chiral limit only) between CSB and charged PC phenomena in the leading order of the large-$N_{c}$ approximation. Moreover, there is also the so-called constrained ${\cal D}_M$ (\ref{pseudo3}) duality of the NJL$_4$ model phase diagram, which is valid even in the physical point \cite{Khunjua:2018sro}. So the significant regions (at $\nu_I,\mu_5\gtrapprox m_\pi/2$) of the particular ($\mu_I$, $T$)- and  ($\mu_5$, $T$)-phase diagrams  should be dually conjugated to each other with respect to a sequential action of two mappings, ${\cal D}$ and ${\cal D}_M$ (see the discussion at the end of Sec. \ref{IIID}), but the duality ${\cal D}$ in the case of the physical point is only approximate (see in Secs. \ref{physpoint} and \ref{IIIC}), although it is valid with a good precision. Since the particular ($\mu_I$, $T$)- and  ($\mu_5$, $T$)-phase diagrams  in these two approaches agrees, one can conclude that the duality can be observed in the lattice QCD simulations.
And this put the notion of the duality on another level of confidence, for it is observed in terms of the toy (1+1)-dimensional NJL model \cite{2dim}, effective (3+1)-dimensional NJL model \cite{Khunjua:2017mkc,Khunjua:2018sro}, lattice QCD simulations and similar dualities has been observed in the large-$N_{c}$ orbifold equivalences approach.
Comparison to the lattice QCD is important not only due to the fact that it is ab initio method for dealing with QCD, but because it does not make use of, for example, large-$N_{c}$ approximation (as in NJL models or in large $N_{c}$ orbifold equivalences approaches).

The question of catalysis of chiral symmetry breaking by chiral chemical potential, i.e. the growth of $T_c$ vs $\mu_5$, is a rather debated one and there are a number of papers that predicted that (pseudo-)critical temperature decrease with increase of chiral chemical potential $\mu_5$ \cite{Ruggieri,Fukushima2}, as well there are a number of papers that support our results \cite{Yu,Xu,Frasca,Cui}. Different regularization schemes have been applied in \cite{Ruggieri2,Yu,Cui,Farias} and it has been stated that if the right one is used there is no catalysis, but lattice QCD results is probably more trustworthy and it disagrees with them.
But the catalysis of chiral symmetry breaking by chiral chemical potential $\mu_5$ can be established in terms of duality notion, let us elaborate on that. As we have talked about, the  ($\mu_I$, $T$)-phase diagram is well established one and the duality fails only in the region of small isospin and chiral isospin chemical potentials (smaller than half of the pion mass), but works quite well for the larger values (see, e.g., in Figs 2, 3). But at the ($\mu_I$, $T$)-phase diagram in the region of isospin chemical potential larger than half of the pion mass the critical temperature increases when $\mu_{I }$ is raised and the duality here is a good approximation, so the critical temperature at the duality conjugated  ($\mu_5$, $T$)-phase diagram should increase with rising of $\mu_{5}$ as well (see at the end of Sec. \ref{IIID}).

Let us summarize the core results of our paper.

\begin{itemize}

\item It has been also demonstrated that the duality correspondence between CSB
and charged PC phenomena observed in Refs. \cite{Khunjua:2017mkc,Khunjua:2018sro} in the chiral limit (where it was exact) is a very good approximate symmetry of the phase diagram even in the physical point in the framework of the NJL$_{4}$ model in the leading order of the large-$N_c$ approximation (see in Sec. \ref{physpoint}). And it stays a very instructive feature of the phase diagram that can be used in different situations.

\item It has been also shown that temperature does not spoil the duality correspondence between CSB and charged PC phenomena
and it stays exact at finite temperature in the chiral limit (see the comment at the end of Sec. \ref{T})  and it is a good approximate symmetry at the physical point (see in Sec. \ref{IIIC}).  

\item We have shown that there is a huge PC${\rm _d}$ phase region in the phase portrait of the model (1) promoted by $\nu_5$ even in the physical point (see, e.g., in Fig. 3). 
And it has been revealed that PC${\rm _d}$ phase can exist at rather large temperatures up to even about 100 MeV (see in Fig. 8). 

\item The particular cross sections of the obtained phase portraits are in qualitative accordance with the recent lattice simulations \cite{Gupta,Braguta:2015zta,Braguta:2015owi}. And it has been established that lattice QCD results support the existence of the duality.

\item The rise of pseudo-critical temperature with increase of chiral chemical potential $\mu_{5}$ has been established in terms of duality notion and the well explored results of lattice QCD and different approaches on phase structure of isotopicaly imbalanced quark matter.  It gives additional argument in favour of this behaviour of the pseudo-critical temperature and it is of importance because, although the lattice results confirming this behaviour are conclusive, the pion mass that is used in these simulations is still quite high and well above the physical pion mass and our results are made at the physical point with the right value of the pion mass. 

\end{itemize}

The central result of our paper is the fact that there is an approximate duality that is supported by recent lattice results. Two different lattice simulations that are completely different and are not connected at the first sight are in reality dual to each other.
Moreover, the logic can be reversed and we can predict the increase of pseudo-critical temperature with rising of chiral chemical potential, the much debated effect recently, just by the duality notion.

This work is intended to generalize and  refine 
the previously obtained results of Refs. \cite{Khunjua:2017mkc,Khunjua:2018sro} to a more physically motivated situation (physical point and finite temperatures). These generalizations require much more computing resources and technically is rather challenging, but it pays off when you can compare the results with lattice QCD and it supports them.
Moreover, we hope that our results might shed new light
on phase structure of dense quark matter with isotopic and chiral imbalance and hence could be of interest in the context of the heavy ion collision experiments and neutron stars interiors.

\section{Acknowledgement}
The authors are thankful to Igor Shovkovy for the idea of the possibility of generation of chiral isospin imbalance in dense matter due to chiral separation effect.

\appendix{}

\section{Generation of nonzero chiral isospin charge in dense quark matter}
\label{ApB}

Let us suppose, for simplicity, that dense quark matter consists of two massless $u$ and $d$ quarks, whose chemical potentials, $\mu_u=\mu+\nu$ and $\mu_d=\mu-\nu$ (see the notations adopted just after the Eq. (\ref{2})), are positive. Moreover, we suppose also that quarks do not interact, and there is an external magnetic field $\vec B=(0,0,B)$ directed along $z$ axis. In this case in the equilibrium state of quark matter there is a nonzero and nondissipative axial current
\begin{eqnarray}
\vec j_{5f}\equiv\vev{\bar q_f\vec \gamma\gamma^5 q_f}=\frac{Q_f\mu_f\vec B}{2\pi^2}\label{B1}
\end{eqnarray}
for each quark flavor $f=u,d$. In Eq. (\ref{B1})  $Q_f$ is an electric charge of the quark flavor $f$, i.e. $Q_u=2/3$, $Q_d=-1/3$. In this case it is not difficult to conclude from Eq. (\ref{B1}) that axial currents of $u$ and $d$ quarks are opposite in their directions.
Since $\vec j_{5f}=\vev{\bar q_{fR}\vec \gamma q_{fR}}-\vev{\bar q_{fL}\vec \gamma q_{fL}}$, where
\begin{eqnarray}
q_{fR}=\frac{1+\gamma^5}{2}q_f,~~~ q_{fL}=\frac{1-\gamma^5}{2}q_f,\label{B2}
\end{eqnarray}
we see from Eq. (\ref{B1}) that left- and right-handed quarks of each flavor $f=u,d$ moves in opposite directions of the $z$ axis. As a result, a spatial separation of quark chiralities for each flavor $f$ occurs. It is the so-called chiral separation effect \cite{Metlitski}. In other words, one can say that in the upper half of the three-dimensional space, i.e. at $z>0$,  the density, e.g., $n_{uR}\equiv\vev{\bar q_{uR}\gamma^0 q_{uR}}$ of the right-handed $u$ quarks is greater than the density $n_{uL}\equiv\vev{\bar q_{uL}\gamma^0 q_{uL}}$ of the left-handed $u$ quarks. Hence, in this case we have at $z>0$ the positive values of the chiral charge density $n_{u5}\equiv n_{uR}-n_{uL}$ for $u$ quarks. (It is evident that at $z<0$ the chiral charge of $u$ quarks is negative.)

On the contrary, since the axial current $\vec j_{5d}$ of $d$ quarks differs in its direction from the axial current $\vec j_{5u}$ of $u$ quarks, one can see that in this case at $z>0$ (at $z<0$) the density $n_{d5}$ of the chiral charge of $d$ quarks is negative (positive). Consequently, we have at $z>0$ the positive values of the quantity  $n_{I5}\equiv n_{u5}-n_{d5}$, which is the ground state expectation value of the density operator for the chiral isospin charge (it is defined in Eq. (\ref{2004})). Whereas at $z<0$ the chiral isospin charge is negative.

In summary, we can say that in dense quark medium under the influence of a strong magnetic field (as an example we can mention  neutral stars), regions with a nonzero chiral isospin charge $n_{I5}$ might appear. Therefore physical processes inside these regions can be described, e.g., in the framework of the Lagrangians of the form (\ref{40}), containing chiral isospin chemical potential $\mu_{I5}$.

\section{Calculation of roots of $P_{\pm}(\eta)$}
\label{ApA}

In this appendix it will be shown how to get roots of the following quartic equation (general quartic equation could be reduced to the one of this form)
\begin{eqnarray}
P_+(\eta)\equiv \eta^4-2a\eta^2+b\eta+c=0.\label{A1}
\end{eqnarray}
The coefficients $a,b,c$ in Eq. (\ref{A1}) are given by the relations (\ref{10}). First, we represent the polynomial on the left-hand side of this equation as the product of two quadratic polynomials,
\begin{eqnarray}
(\eta^2+r\eta+q)(\eta^2-r\eta+s)=0,\label{A2}
\end{eqnarray}
where
$$
-r^{2}+q+s=-2a,\quad qs=c,\quad rs-rq=b.
$$
It follows from these relations that
\begin{eqnarray}
q=\frac{1}{2} \left(-2 a+r^2-\frac{b}{r}\right),~~~s=\frac{1}{2} \left(-2 a+r^2+\frac{b}{r}\right).\label{A3}
\end{eqnarray}
Substituting Eq. (\ref{A3}) into Eq. (\ref{A2}), one gets
that $r=\sqrt{R}$, where $R$ is one of the solutions of the following cubic equation
\begin{equation}
X^{3}+AX=BX^{2}+C,
\label{cub13}
\end{equation}
where we used notations $A, B, C$ that are given by
$$
A=16(\Delta ^2 {\nu _5}^2+ \nu ^2{ \nu _5}^2+ \nu ^2 M^2+ p^2 \left(\nu ^2+{\nu _5}^2\right)),~~~B=4a,~~~C=b^2.
$$
All three solutions of the cubic equation (\ref{cub13}) are
\begin{equation}
R_{1,2,3}=\frac{1}{3} \left(4 a+\frac{L}{\sqrt[3]{J}}+\sqrt[3]{J}\right),\label{A5}
\end{equation}
where
$$
J=\frac{1}{2}(K+i\sqrt{4 L^3-K^2}),~~K=128 a^3-36 a A+27 b^2,~~L=-3A+16 a^2,
$$
and $ \sqrt[3]{J}$ in Eq. (\ref{A5}) means each of three possible complex valued roots. There is a determinant $D\equiv 4L^{3}-K^{2}>0$ of the equation (\ref{cub13}) that can tell us the structure of roots $R_{1,2,3}$. Namely,
if $D>0$ then all roots $R_i$ are real and different, if $D=0$ all roots are real and at least two are equal. Finally, if $D<0$ then one root is real and two are complex conjugate. So, there is always a real solution of Eq. (\ref{cub13}). In numerical simulations it is more handy to work with real solution and it is always possible to choose one.
There is a procedure that, depending on values of parameters, chooses a real solution, but it is quite lengthy so we will not present it here.

And when one has found $r$, the roots of Eq. (\ref{A1}) has the following form
\begin{equation}
\eta_{1}=\frac{1}{2} \left(-\sqrt{r^2-4 q}-r\right),
\eta_{2}=\frac{1}{2} \left(\sqrt{r^2-4 q}-r\right),
\eta_{3}=\frac{1}{2} \left(r-\sqrt{r^2-4 s}\right),
\eta_{4}=\frac{1}{2} \left(r+\sqrt{r^2-4 s}\right). \label{A6}
\end{equation}
The roots $\eta_{5,6,7,8}$ of the equation $P_-\equiv\eta^4-2a\eta^2-b\eta+c=0$ can be obtained by changing $b\to-b$ in Eq. (\ref{A1}) (or $q\leftrightarrow s$ in Eq. (\ref{A6}) with $r$ unchanged). So, we have
 $$
\eta_{5}=-\eta_{4},\,\eta_{6}=-\eta_{3},\,\eta_{7}=-\eta_{2},\,
\eta_{8}=-\eta_{1}.
$$

\end{document}